\documentclass[%
 aip,
 amsmath,amssymb,
 reprint,%
]{revtex4-1}

\usepackage{graphicx}
\usepackage{dcolumn}
\usepackage{bm}
\usepackage{hyperref}
\usepackage[utf8]{inputenc}
\usepackage[T1]{fontenc}
\usepackage{amsmath}
\usepackage{braket}
\usepackage[utf8]{inputenc}

\usepackage{bbold}
\usepackage{subcaption}

\preprint{AIP/123-QED}
\begin{document}

\title{Optimization of the time-multiplexed SPDC source at 900-950 nm range}

\author{V.\, O.~Gotovtsev} \email{vladgotovtsev@mail.ru}
\affiliation{Quantum Technology Centre and Faculty of Physics, M.V. Lomonosov Moscow State University, 1 Leninskie Gory Street, Moscow 119991, Russian Federation }
\author{I.\,V.~Dyakonov}
\affiliation{Quantum Technology Centre and Faculty of Physics, M.V. Lomonosov Moscow State University, 1 Leninskie Gory Street, Moscow 119991, Russian Federation }
\affiliation{Russian Quantum Center, Bolshoy bul’var 30 building 1, Moscow 121205, Russian Federation }
\author{O.\,V.~Borzenkova}
\affiliation{Quantum Technology Centre and Faculty of Physics, M.V. Lomonosov Moscow State University, 1 Leninskie Gory Street, Moscow 119991, Russian Federation }
\affiliation{Russian Quantum Center, Bolshoy bul’var 30 building 1, Moscow 121205, Russian Federation }
\author{K.\, A.~Taratorin} 
\affiliation{Quantum Technology Centre and Faculty of Physics, M.V. Lomonosov Moscow State University, 1 Leninskie Gory Street, Moscow 119991, Russian Federation }
\author{T.\, B.~Dugarnimaev} 
\affiliation{High school of economics, 11 Pokrovsky Bulvar, Moscow, Russian Federation }
\author{A.\, A.~Korneev} 
\affiliation{Quantum Technology Centre and Faculty of Physics, M.V. Lomonosov Moscow State University, 1 Leninskie Gory Street, Moscow 119991, Russian Federation }
\author{S.\, P.~Kulik} 
\affiliation{Quantum Technology Centre and Faculty of Physics, M.V. Lomonosov Moscow State University, 1 Leninskie Gory Street, Moscow 119991, Russian Federation }
\author{S.\, S.~Straupe}
\affiliation{Quantum Technology Centre and Faculty of Physics, M.V. Lomonosov Moscow State University, 1 Leninskie Gory Street, Moscow 119991, Russian Federation }
\affiliation{Russian Quantum Center, Bolshoy bul’var 30 building 1, Moscow 121205, Russian Federation }

\date{\today}

\begin{abstract}
    In the field of quantum technology, single photons have emerged as a pivotal resource, prompting the development of heralded single photon sources (HSPS) with enhanced generation probability. The majority of such sources are based on spontaneous parametric down-conversion (SPDC), but they exhibit a low single photon generation probability. The multiplexing principle \cite{pittman2002single} has been proposed as a solution to this problem. This paper presents a demonstration of a time-multiplexed HSPS based on the SPDC process, including accurate calculations and modeling of key source characteristics, specifically purity and heralding efficiency. Furthermore, the paper provides an analysis and approximation of the probability of a single photon post-application of time multiplexing.
\end{abstract}

\maketitle
\section{Introduction}

Single photons are indispensable building blocks for modern quantum technologies. They can distribute information across quantum communication networks \cite{monroe2002quantum, northup2014quantum}, serve as carriers of quantum states in memories and processors \cite{afzelius2015quantum, wang2019efficient}, enhance sensitivity in metrology and sensing applications \cite{pirandola2018advances, lawrie2019quantum, degen2017quantum}, and provide a robust testbed for new quantum protocols and fundamental physics \cite{flamini2018photonic}. Because of this versatility, the development of reliable single-photon sources remains a central challenge in photonic quantum science.

Two broad strategies for generating single photons have been pursued. The first involves single quantum emitters—neutral atoms \cite{hennrich2004photon, hijlkema2007single}, ions \cite{keller2004continuous, barros2009deterministic, maurer2004single}, molecules \cite{kiraz2005indistinguishable, fleury2000nonclassical}, quantum dots \cite{senellart2017high, arakawa2020progress}, and color centers in diamond \cite{aharonovich2011diamond, aharonovich2016solid}. These systems promise deterministic single-photon emission with high indistinguishability, but they require demanding experimental conditions, exhibit limited spectral tunability, and face scalability challenges. The second strategy relies on spontaneous nonlinear processes—spontaneous parametric down-conversion (SPDC) \cite{kaneda2015time, pittman2002single, ma2011experimental, magnitskiy2015spdc} and spontaneous four-wave mixing (SFWM) \cite{goldschmidt2008spectrally, fiorentino2002all, li2005optical}. These processes are intrinsically probabilistic \cite{christ2012limits}, but they are technologically simple, robust against environmental noise, and compatible with integrated photonics platforms.

A key advantage of SPDC and SFWM is that they generate photon pairs. Detection of one photon (idler) heralds the presence of its partner (signal), enabling heralded single-photon sources (HSPS). Heralding provides precise information about the emission time and spatial mode, allowing photons to be routed, synchronized, or stored. This has made HSPS one of the most widely used sources in photonic quantum technologies. However, because the underlying nonlinear process is probabilistic, HSPS must balance generation probability, heralding efficiency, and suppression of multiphoton events.

The heralding mechanism provides crucial information about the timing and location of photon emission, thereby overcoming the inherent probabilistic nature of photon generation. Once this information has been obtained, the photon can be selected, stored, and redirected to be emitted at a desired time and in a specific optical mode. When multiple probabilistic HSPSs are available, the multiplexing principle can be used to integrate an array of probabilistic sources into a near-deterministic photon source \cite{joshi2018frequency, jeffrey2004towards}.

A nonlinear HSPS is a convenient photon source that offers flexible wavelength tuning over a broad range of highly indistinguishable photons and is easily integrated with high yield with current photonic technology. The probability of HSPS generation is interrelated with the purity of the source due to a seamless transition to the squeezed vacuum generation regime \cite{}.

In order to simultaneously generate $k$ photon pairs using a single pump pulse, the probability of this process is governed by the relationship
\begin{equation} \label{eq:P_k}
    P(k) = \frac{\mu^k}{(\mu+1)^{k+1}},
\end{equation}
where $\mu$ denotes the average number of pairs per pump pulse. The maximum probability of generating a single-photon pair ($k=1$) is attained at $\mu=1$, corresponding to $25\%$ \cite{kaneda2015time,kaneda2019high}. However, this configuration also entails a considerable probability of either generating zero-photon pairs or multiple-photon pairs. To mitigate the generation of multiphoton pairs, $\mu$ is often deliberately kept low in many experiments. Unfortunately, this approach concomitantly results in a reduced rate of single coincidences.

Over the past two decades, HSPS have been realized in a variety of physical systems. Bulk-crystal SPDC sources have demonstrated high state purity and low multi-photon contamination, but collection efficiencies are limited by spatial mode mismatch. Fiber-based SFWM sources offer compatibility with telecommunication wavelengths and long-distance transmission, though they suffer from Raman noise and propagation losses. Cold atomic ensembles have provided narrowband heralded photons well matched to atomic transitions, enabling strong light–matter coupling, but require complex experimental infrastructure. Heralding has also been explored with quantum dots, combining triggered emission with auxiliary detection, though device-to-device variability and spectral diffusion remain significant obstacles. Across these platforms, key performance metrics include heralding efficiencies exceeding $80\%$ \cite{pittman2002single, ma2011experimental}, second-order correlations $g^{(2)}(0)<0.01$ \cite{kaneda2015time}, and spectral purities approaching $90\%$ with proper filtering or mode engineering. Yet in all cases, practical use is constrained by the trade-off between brightness, purity, and multiphoton suppression.

To overcome the probabilistic nature of nonlinear photon generation, multiplexing schemes have been developed. These approaches combine multiple HSPS into a single output, effectively increasing the probability of delivering a single photon on demand. Spatial multiplexing uses arrays of sources connected by active switching networks \cite{jeffrey2004towards}; temporal multiplexing employs fast optical switches and fiber delay lines to redirect heralded photons to a common output channel \cite{kaneda2015time}; and frequency multiplexing exploits distinct spectral modes within the same nonlinear process \cite{joshi2018frequency}. More recently, chip-scale platforms have integrated such multiplexing schemes to enhance scalability, though bulk and fiber-based implementations remain highly relevant for laboratory and field applications.

In this work, we investigate the prospects of photon multiplexing using a single-photon source based on spontaneous parametric down-conversion (SPDC) at $900$-$950\;nm$ range. We demonstrate how multiplexing can enhance the probability of generating a photon within a well-defined temporal window from 0.005 to 0.008. To quantify this effect, we introduce the concept of heralding probability \ref{sec:HE_and_purity}, which we define in terms of the joint spectral amplitude \ref{sec:JSA} of the photon pairs. Furthermore, we analyze how these characteristics, together with the photon purity, can be optimized to achieve high-performance single-photon generation with SPDC sources.

\section{Joint spectral amplitude}  \label{sec:JSA}

To analyze the performance of the SPDC source in detail, we make use of the joint spectral amplitude (JSA). The JSA fully characterizes the two-photon state in the frequency domain by describing the spectral correlations between signal and idler photons. It provides a convenient framework for evaluating both the heralding probability and the purity of the heralded single-photon state. In what follows, we introduce the formal definition of the JSA and discuss how it can be employed to assess the key performance parameters of SPDC-based sources.

The biphoton state generated by an SPDC crystal has the following wave function\cite{bennink2010optimal}:
\begin{equation} \label{eq:PSI_SPDC}
    |\Psi_{SPDC} \rangle = -i\int_0^\infty d\omega_s d\omega_i \psi(\omega_s,\omega_i) \hat{a}_{\omega_s}^\dagger\hat{a}_{\omega_i}^\dagger |vac\rangle,
\end{equation}
where $\hat{a}^\dagger_{\omega_{s/i}}$ - operator that creates a photon in signal/idler mode and
\begin{equation} \label{eq:JSA_general}
    \psi(\omega_s,\omega_i) = \sqrt{\frac{2\pi^2\hbar N_p}{\varepsilon_0\lambda_p\lambda_s\lambda_i}}s(\omega_p)O(\omega_s,\omega_i)
\end{equation}
is the JSA, which contains all the information about the correlation between the signal and idler photons. Here, $N_p$ -- the mean number of pump photons, $\varepsilon_0$ -- the vacuum permittivity, $s(\omega_p)=exp(-\frac{(\omega_p-\omega_{p0})^2}{4\sigma_p^2})$ -- the pump spectral amplitude, $O(\omega_s,\omega_i)$ -- the spatial overlap of the pump, signal, and idler modes in the medium, which generalizes the $sinc$ phase matching function encountered in plane-wave treatments of SPDC. In the case of Gaussian beams propagating along the z-axis, the function of spatial overlap is defined as:
\begin{multline}\label{eq:Overlap_func}
        O(\omega_s,\omega_i) =\\
            \sqrt{\frac{8\epsilon}{\pi}}\chi_{eff}^{(2)}w_pw_sw_i\int_{-L/2}^{L/2}\frac{exp[i(\Delta k+mK)z]}{q_s^*q_i^*+q_pq_s^*+q_pq_i^*}dz, 
\end{multline}
where $\epsilon$ -- an efficiency factor, $\chi_{eff}^{(2)}$ -- the effective nonlinear coefficient, $w$ -- the waist size, $\Delta k = k_p-k_s-k_i$, $K=2\pi/\Lambda$, $q=w^2+2iz/k$. In order to facilitate the manipulation of \eqref{eq:Overlap_func}, it is necessary to introduce dimensionless quantities. The primary quantities of interest are the phase mismatch
\begin{equation} \label{eq:phase_mismatch}
    \Phi = (\Delta k+mK)L,
\end{equation}
which is responsible for momentum conservation, and the focal parameters
\begin{equation} \label{eq:focal_par}
    \xi_j = \frac{L}{k_jw_j^2},
\end{equation}
which describe the degree ($\xi_j \gg 1$ - strongly, $\xi_j \ll 1$ - weakly) of confinement of field j ($j = p,s,i$) compared to the crystal length. With this definition at hand, the original equation \eqref{eq:JSA_general} can be rewritten using the entered values as follows:
\begin{eqnarray} \label{eq:JSA}
    \psi(\omega_s,\omega_i) = &&\sqrt{\frac{8\pi^2\epsilon\hbar n_sn_iN_pL}{\varepsilon_0n_p}}\frac{\chi_{eff}^{(2)}}{\lambda_s\lambda_i}\frac{s(\omega_p)}{\sqrt{A_+B_+}} \nonumber\\
    &&\times \int_{-1}^1\frac{\sqrt{\xi}exp(i\Phi l/2)}{1-i\xi l - C\xi^2l^2}dl.
\end{eqnarray}
In this instance, the following designations are employed for the purpose of simplification:
\begin{equation} \label{eq:A_+}
    A_+=1+\frac{k_s}{k_p}\frac{\xi_s}{\xi_p}+\frac{k_i}{k_p}\frac{\xi_i}{\xi_p};
\end{equation}
\begin{equation} \label{eq:B_+}
    B_+=\left(1-\frac{\Delta k}{k_p}\right)\left(1+\frac{k_s+\Delta k}{k_p-\Delta k}\frac{\xi_p}{\xi_s}+\frac{k_i+\Delta k}{k_p-\Delta k}\frac{\xi_p}{\xi_i}\right);
\end{equation}
\begin{equation} \label{eq:C}
    C = \frac{\Delta k}{k_p}\frac{\xi_p^2}{\xi_s\xi_i}\frac{A_+}{B_+^2};
\end{equation}
\begin{equation} \label{eq:agg_focal_par}
    \xi = \frac{B_+}{A_+}\frac{\xi_s\xi_i}{\xi_p}.
\end{equation}

To acquire JSA, it is essential to compute the optimal parameters that fulfill the phase matching condition, as referenced in $\Phi = 0$.
In this study, a ppKTP crystal $2.0\;mm\,(x) \times 2.0\;mm\,(y) \times 1.0\;mm\,(z)$ characterized by a domain period of 29.4 µm is employed, with the pump beam directed along the x-axis. Collinear second-type spontaneous parametric down-conversion (SPDC) is observed utilizing a 465 nm pump, which is horizontally polarized and possesses a full width at half maximum of 12 nm. The wavelengths corresponding to the signal and idler photons are 925 nm (vertically polarized) and 935 nm (horizontally polarized), respectively.

According to\cite{kovlakov2017spatial} for generation in single mode, optimal beam waist is determined by the condition in the length of crystal fits two Rayleigh length and we get $w_j = \sqrt{L/k_j}$. Such a way optimal focal parameters will be equal 1.

\begin{figure}[!h]
    \centering
    \includegraphics[width=1\linewidth]{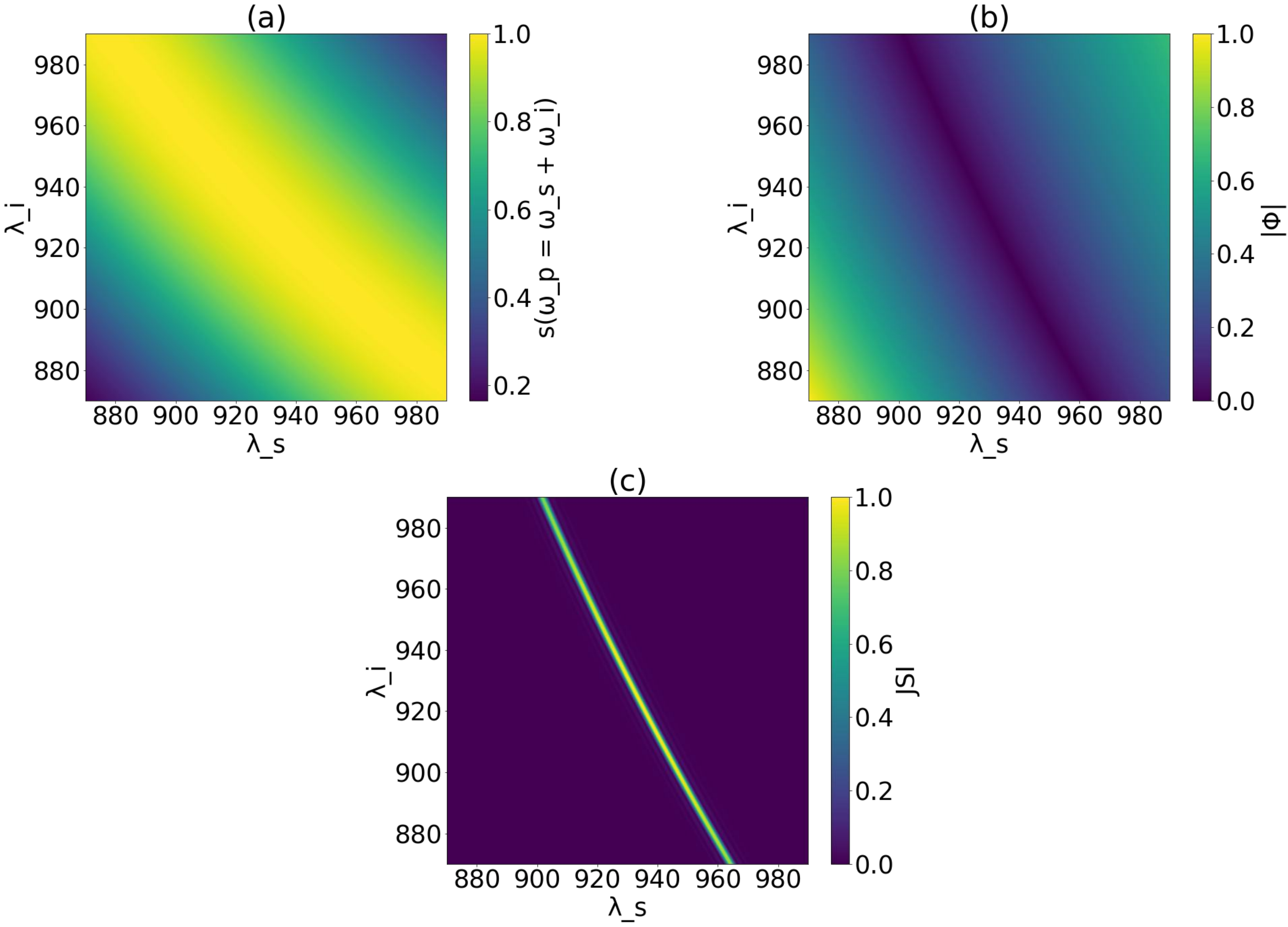}
    \caption{(a) The pump spectral envelope function -- $s(\omega_p=\omega_s+\omega_i)$; (b) The phase mismatch function module -- $|\Phi|$; (c) Joint spectral intensity -- $|\psi(\omega_s,\omega_i)|^2$}
    \label{fig:JSI}
\end{figure}

Taking into account the characteristics of the crystal, the phase matching settings and the central pump frequency, the calculation of $\psi(\omega_s, \omega_i)$ is performed. Figure \ref{fig:JSI} (a) illustrates the properties of the pump $s(\omega_p=\omega_s+\omega_i)$. Referring to Figure \ref{fig:JSI}(b), it is possible to determine the wavelengths that satisfy the phase matching condition $\Phi=0$. The joint spectral intensity function (JSI), presented in Figure \ref{fig:JSI}(c), includes the squared modulus of the product of the aforementioned functions and encapsulates the characteristics of the simulated source.

\section{Heralding efficiency and Purity} \label{sec:HE_and_purity}

Having established the JSA framework and its dependence on source parameters, we now consider the heralding efficiency (HE), which is defined as the probability of detecting a photon when its paired photon has already been detected \cite{kaneda2016heralded}. This can be mathematically represented as: $\eta_{s(i)} = \frac{P_{si}}{P_{i(s)}t_{s(i)}d_{s(i)}}$, where $P_{s(i)}$ denotes the single-photon collection probability in the signal (idler) channel, $P_{si}$ signifies the pair collection probability,$t_{s(i)}$ and $d_{s(i)}$ represents the total transmission and the detection efficiency of the signal (idler) mode.

Utilising the JSA \eqref{eq:JSA} function, the pair collection probability can be determined by\cite{bennink2010optimal}:
\begin{equation} \label{eq:P_si_general}
    P_{si} = \int |\psi(\omega_s,\omega_i)|^2d\omega_sd\omega_i
\end{equation}
Given that an exact integration of \eqref{eq:JSA} cannot be readily carried out, a more feasible approach is to employ controlled approximations. As a first step, we introduce these approximations and assess the extent to which they remain valid:
\begin{eqnarray} \label{eq:approx_1} 
    &&C \approx 0 \\ 
    &&A_+,B_+,\xi \approx const \label{eq:approx_2}
\end{eqnarray}

The fairness of this simplifications will be discussed in Appendix A.

\begin{figure}[!h]
  \centering
  \includegraphics[width=1\linewidth]{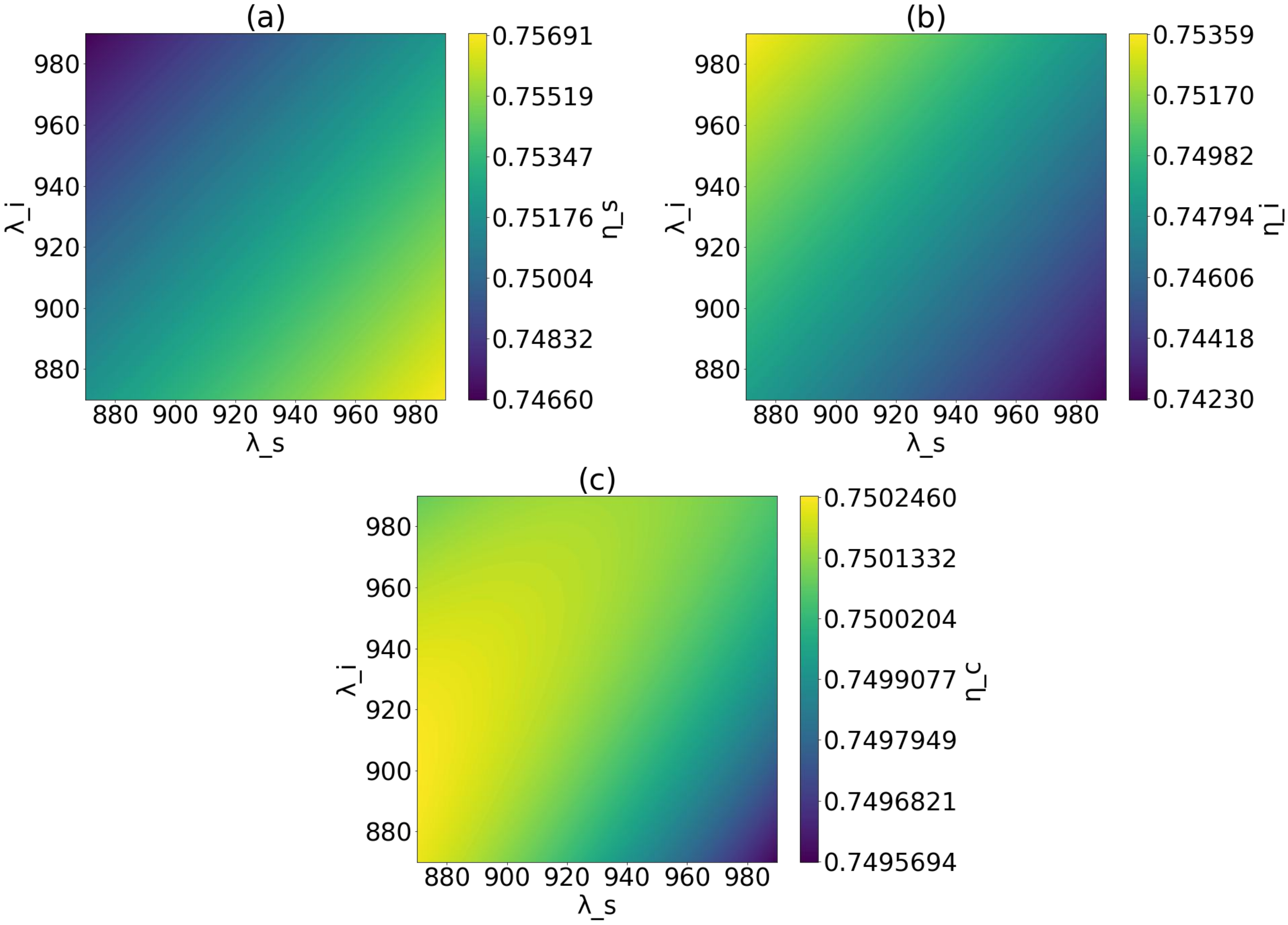}
  \caption{Heralding efficiency dependence of wavelength (a) -- signal heralding efficiency, (b) -- idler heralding efficiency, (c) -- symmetric heralding efficiency \label{fig:heralding efficiency}}
\end{figure}

Utilizing approximations \eqref{eq:approx_1},\eqref{eq:approx_2}, equation \eqref{eq:P_si_general} can be rewritten in a simplified form:
\begin{equation}\label{eq:P_si}
    P_{si} =\frac{64\pi^3\hbar c\epsilon n_sn_i}{\varepsilon_0n_p|n_s'-n_i'|}\Big(\frac{\chi_{eff}^{(2)}}{\lambda_s\lambda_i}\Big)^2\frac{arctan(\xi)}{A_+B_+}N_p,
\end{equation}
where $n_j'=c\partial k_j/\partial w_j$ -- the group index of mode j. Equations of a similar nature has the single-photon collection probability in signal(idler) mode:
\begin{equation}\label{eq:P_s}
    P_{s(i)} =\frac{64\pi^3\hbar c\epsilon n_sn_i}{\varepsilon_0n_p|n_s'-n_i'|} \Big(\frac{\chi_{eff}^{(2)}}{\lambda_s\lambda_i}\Big)^2\frac{arctan(\frac{B_{s(i)}}{A_{s(i)}}\xi_{s(i)})}{A_{s(i)}B_{s(i)}}N_p,
\end{equation}
where $A_{s(i)}=2\sqrt{\Big(1+\frac{k_{s(i)}}{k_p}\frac{\xi_{s(i)}}{\xi_p}\Big)\frac{k_{i(s)}}{k_p}}$, $B_{s(i)}=2(1-\frac{\Delta k}{k_p})\sqrt{(1+\frac{k_{s(i)}+\Delta k}{k_p-\Delta k}\frac{\xi_p}{\xi_{s(i)}})\frac{k_{i(s)}+\Delta k}{k_p-\Delta k}}$. The equation for signal/idler heralding efficiency is then given by (see Figure \ref{fig:heralding efficiency}(a,b)):
\begin{equation}\label{eq:eta_s(i)}
    \eta_{s(i)} = \frac{arctan(\xi)A_{s(i)}B_{s(i)}}{A_+B_+arctan(\frac{B_{s(i)}}{A_{s(i)}}\xi_{s(i)})}
\end{equation}
Additionally, the symmetric heralding efficiency is considered (see Figure \ref{fig:heralding efficiency}(c)):
\begin{equation}\label{eq:eta_c}
    \eta_c = \frac{P_{si}}{\sqrt{P_sP_i}} = \frac{\sqrt{A_sA_iB_sB_i}arctan(\xi)}{A_+B_+\sqrt{arctan(\frac{B_s}{A_s}\xi_s)arctan(\frac{B_i}{A_i}\xi_i)}}
\end{equation}
Utilizing the parameters of our crystal and equations (\ref{eq:eta_s(i)},\ref{eq:eta_c}), we obtain a heralding efficiency value of 0.75 for the entire wavelength range under consideration (see Figure \ref{fig:heralding efficiency}).

\begin{figure}[!h]
  \centering
  \includegraphics[width=1\linewidth]{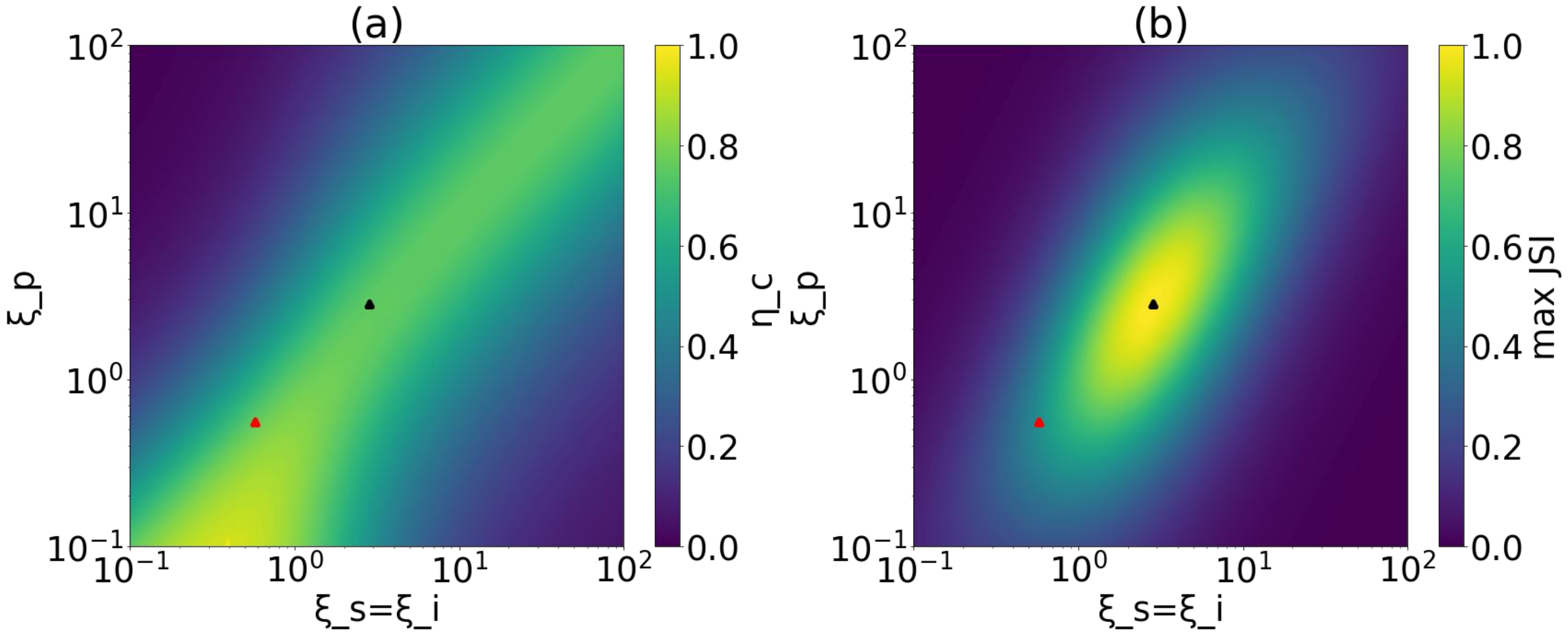}
  \caption{(a) Symmetric heralding efficiency dependence of pump and signal/idler focal parameters. (b) Maximum JSI value depending on focal parameters. Red dot corresponds to $\xi_p=\xi_s=\xi_i\approx 0.55$ that we got in our model. The red marker corresponds to the optimum values $\xi_p=\xi_s=\xi_i\approx 2.84$ \label{fig:n_c_max_JSI}}
\end{figure}

As demonstrated in Figure \ref{fig:heralding efficiency}, the heralding efficiency demonstrates a negligible dependence on wavelength. Consequently, to improve heralding efficiency, it is imperative to consider the dependence on focal parameters \eqref{eq:focal_par}. As demonstrated in Figure \ref{fig:n_c_max_JSI}(a), the optimal focal parameters for the efficiency of symmetric heralding are within the range $\xi_p\lesssim 0.1$, $0.04 \lesssim \xi_s=\xi_i\lesssim 0.5$. Implementation of these parameters is possible with the use of an ultra-short crystal ($L < 0.4\:mm$) or using a wider pump waist size. However, it should be noted that a reduction in the focal parameter results in a weak field focus relative to the crystal length, leading to a decline in the photon generation efficiency. As illustrated in Figure \ref{fig:n_c_max_JSI}(b), the maximum attainable value of JSI is dependent on the focal parameters, with optimal generation occurring at values within the range of $1<\xi_{p,s,i}<10$, reaching a maximum at $\xi_p=\xi_s=\xi_i=2.84$. It is therefore necessary to find a compromise between the efficiency of heralding and generation rate.

Another critical factor influencing the quality of a photon source is the purity of the generated state. The utilisation of spectral filters is frequently employed to enhance purity; however, this practice concomitantly results in a reduction in prediction efficiency. The following examination will address the manner in which spectral filters influence these parameters.

The purity $P$ can be calculated using the established connection with the Schmidt number (equation \eqref{eq:Purity}). By performing a singular value decomposition (SVD) on the JSA, it is possible to determine the Schmidt number $K$ and subsequently derive its inverse, which represents the purity \cite{zielnicki2018joint}:
\begin{eqnarray}\label{eq:Purity}
    &&|\psi> = \sum_i \sqrt{\lambda_i} u_i(\omega_s)v_i(\omega_s),\nonumber \\
    &&K = \frac{1}{\sum_i \lambda_i^2}, \\
    &&P = \frac{1}{K} \nonumber
\end{eqnarray}

\begin{figure}[!h]
  \centering
  \includegraphics[width=1\linewidth]{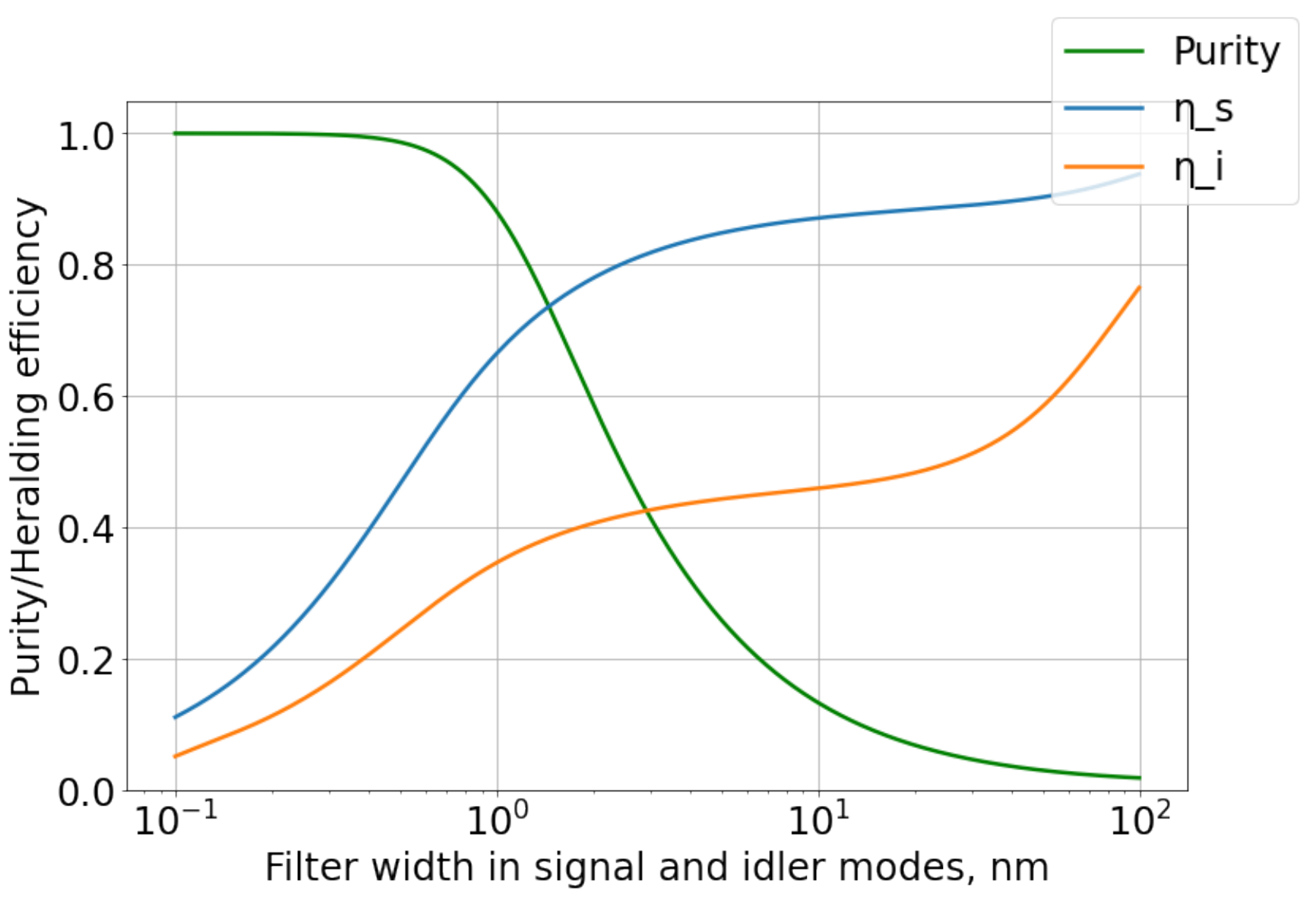}
  \caption{Purity (green line) and normalized value of heralding efficiency (blue and orange lines) versus filters width\label{fig:Purity}}
\end{figure}

The configuration of the JSA is contingent upon the quantity of Schmidt modes within the system. In circumstances where the number of modes is reduced, the JSA profile has a propensity to adopt a circular form. The application of filters to the signal and idler modes approximates the JSA to a more circular configuration, thereby reducing the number of Schmidt modes and enhancing the purity value. In the case presented in Figure \ref{fig:JSI}(c), the JSA spectrum is extensive, resulting in a predicted low purity. It is evident from calculation \eqref{eq:Purity} that the value obtained is $P\approx0.014$. As illustrated by the green line in Figure \ref{fig:Purity}, there is a demonstrable correlation between purity and filter widths in signal and idler mode, with the purity value approaching its maximum as the filter width decreases. However, it should be noted that the use of spectral filters has been shown (see blue and orange lines on Figure \ref{fig:Purity}) to decrease heralding efficiency. In order to achieve a purity of at least $0.9$, it is necessary to employ spectral filters with a bandwidth of approximately $0.93\;nm$. However, this will result in a normalized value of heralding efficiency of $0.65$ and $0.34$ in the signal and idler channel, respectively. In this regard, there is a necessity to explore alternative methods for enhancing the purity, a subject that will be addressed in Appendix B.

\section{Time multiplexing}

\begin{figure}[!h]
  \centering
  \includegraphics[width=1\linewidth]{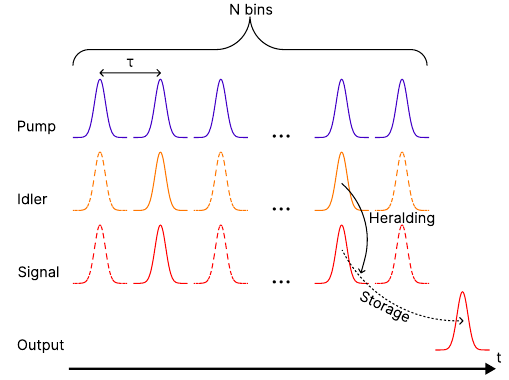}
  \caption{Time multiplexing scheme. We pump nonlinear crystal with a period $\tau$ and probabilistically generate photon pare during different N time bins. With detecting idler photons we put into the time storage the last heralded signal photon and output it after considered N time bins.
  \label{fig:Time multiplexing}}
\end{figure}

A time multiplexing technique involves pumping a crystal with a sequence of N pulses with the objective of observing at least one photon pair. Following the activation of the trigger by idler photon, the signal one is stored in an optical memory and released after a pump pulse series with defined duration (Figure \ref{fig:Time multiplexing}). When considering N multiplexed modes, with idler mode trigger probability p, the probability of getting a heralded photon is equal:
\begin{equation}\label{eq:P_H}
    P_H = 1 - {(1-p)}^N. 
\end{equation}

To calculate the probability of producing a single-photon state after time multiplexing, it is necessary to modify the formula \eqref{eq:P_H} to account for the variable parameters of the experiment. As mentioned above, the probability of generating pairs of k-photons is defined by \eqref{eq:P_k}. In our experiment $\mu\thicksim 10^{-3}$, consequently the probability of generating more than one pair of photons is $P(k>1)\thicksim 10^{-6}$. In view of the aforementioned considerations, the generation of multiple-photon pairs from subsequent analyses can be disregarded. Considering use only heralded photons we should taking into account heralding efficiency in the single-photon pair probability as $P(1)=\frac{\mu}{(\mu+1)^2}\eta_{s(i)}$.

In the observation of $N$ pump pulses, a photon from the last generated pair is directed to a controlled delay line where it is retained until the end of the series. The probability of generating a pair in the $k$-th pulse and no pairs in the subsequent $N-k$ pulses is then given by $P(1)(1-P(1))^{N-k}$. In this instance, the generation of photons prior to the $k$-th pulse is also taken into account. Prior to the insertion of the photon into the controlled delay line, it is first required to enter a delay line, the purpose of which is to allow time for the analysis of all pulses by the control board. Subsequently, the photon enters an optical memory cell, where it remains until the conclusion of the cycle. The probability of the loss of a photon in a single rotation within the memory cell is denoted by $\eta_{sl}$. The probability of preserving the $k$-th photon until the end of the cycle is calculated as $(1-\eta_{sl})^{(N-k+1)}$.

\begin{figure*}
  \centering
  \includegraphics[width=1\linewidth]{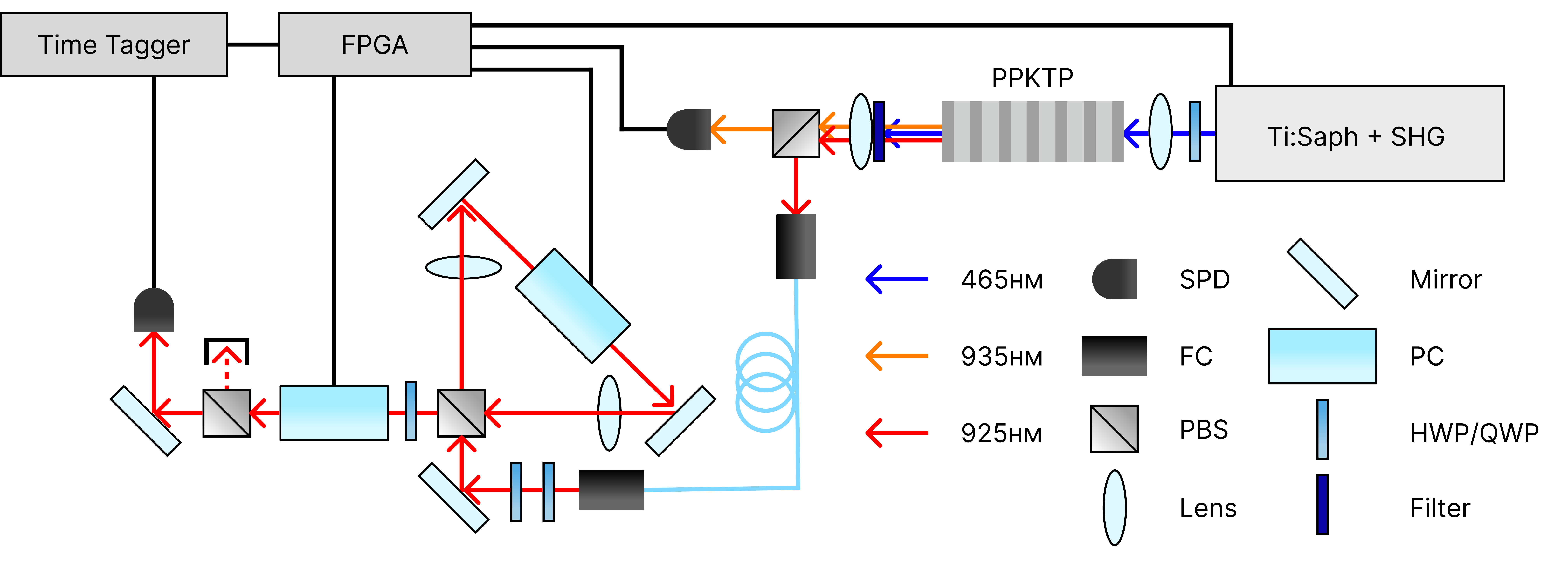}
  \caption{Schematic diagram of our experimental setup. Ti:Saph - titanium-sapphire laser; SHG - second garmonic generator; PPKTP - periodically poled potassium titanyl phosphate; SPD - single photon detector; FC - fiber coupler; PC - Pockels cell; FPGA - field-programmable gate array \label{fig:Experiment}}
\end{figure*}

The total probability of detecting the photon after N multiplexed modes is given by:
\begin{eqnarray}\label{eq:P_multiplexed}
    &&P_{multiplexed} = \nonumber\\
    &&= \sum_{k=1}^N \left( P(1){(1-P(1))}^{N-k}{(1-\eta_{sl})}^{(N-k+1)} \right) = \nonumber\\
    &&= P(1)(1-\eta_{sl})\times \frac{(1-P(1))^N(1-\eta_{sl})^N-1}{(1-P(1))(1-\eta_{sl})-1}.
\end{eqnarray}

\section{Experiment}

In order to construct a single photon source and subsequently analyze its characteristics, the experimental setup depicted in Figure \ref{fig:Experiment} was utilized \cite{kaneda2015time,kaneda2019high}. The experimental setup incorporated a titanium-sapphire femtosecond laser, which generated pulses with a wavelength of $930\; nm$ and a spectral width of $10\;nm$. The duration of these pulses was approximately $100\;fs$, and their period was $13 \;ns$.

After generation the idler photon, exhibiting horizontal polarization, was directed through a polarizing beam splitter (PBS) prior to being transmitted to a detector. The signal obtained from the detector was then processed by a field-programmable gate array (FPGA). Upon detecting a signal indicating the presence of an idler photon in the channel, the FPGA recorded the pulse number within the specified series.

Subsequently, the FPGA selected the most recently detected photon and employed a Pockels cell to store the corresponding signal photon in an optical memory cell. The signal photon remained confined within the optical memory cell until the end of the series, at which point it was released and directed to a detector for further analysis. But Pockels cells have inter limitation for repeated actuation $\sim 65\;ns$. So we can't store first and last five photons during the series. Using FPGA we can make special actuation for the last photon in the series by doubling opening time of Pockels cell. Such a way we should sum from $k=5$ to $k=N-5$ and separately add $k=N$, so equation \eqref{eq:P_multiplexed} will transform into:
\begin{eqnarray}\label{eq:P_multiplexed_work_zone}
    &&P_{multiplexed} = P(1)(1-\eta_{sl}) + \nonumber\\
    &&+P(1)(1-P(1))(1-\eta_{sl})^6\times\nonumber\\
    &&\times \frac{(1-P(1))^{N-10}(1-\eta_{sl})^{N-10}-1}{(1-P(1))(1-\eta_{sl})-1}.
\end{eqnarray}

\begin{figure}[!h]
  \centering
  \includegraphics[width=1\linewidth]{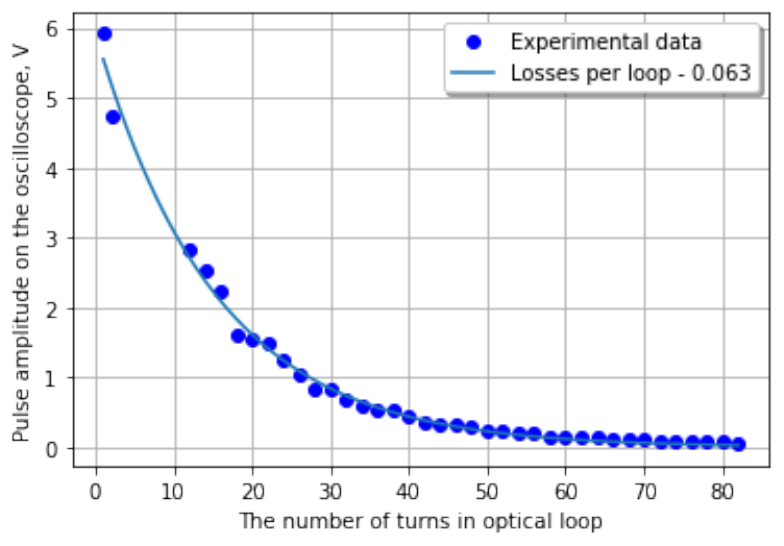}
  \caption{Experimental data of the measured amplitude of the pulse held in the optical memory cell for a controlled number of turns. The blue dots - experimental data. The blue line - an approximation by function $y=a*(1-\eta_{sl})^x$, here $\eta_{sl}$ - losses per loop.\label{fig:Sagnac losses}}
\end{figure}

In order to enhance the efficiency of single-photon generation, it is imperative to minimize losses within the optical memory cell and to improve heralding efficiency. In our experiments, the frequency of single photons in the horizontal mode was measured at $2,422\; MHz$, while in the vertical mode it was recorded at $2,548\; MHz$. Subsequent to the consideration of fiber losses ($\eta_t=0.80$) and detector efficiency ($\eta_{di}=0.70,\: \eta_{ds}=0.67$), the following values for heralding efficiency were obtained: $\eta_{H}=0.338, \eta_{V}=0.339$. In order to optimize the positioning of the optical elements within the memory cell, a single laser pulse was held in the cell for a predetermined number of turns. The resulting signal was then measured using a silicon fixed-gain detector and displayed on an oscilloscope. The graph illustrating the relationship between power loss and the number of turns in the memory cell was subsequently obtained (see Figure \ref{fig:Sagnac losses}).

Consequently, the probability of obtaining a single-photon state will be calculated using the obtained data and a loss value of $0.063$. Utilizing the expression obtained \eqref{eq:P_multiplexed_work_zone} and the experimental data from the setup, the dependence of the photon detection probability on the number of multiplexed modes is plotted.

\begin{figure}[!h]
  \centering
  \includegraphics[width=1\linewidth]{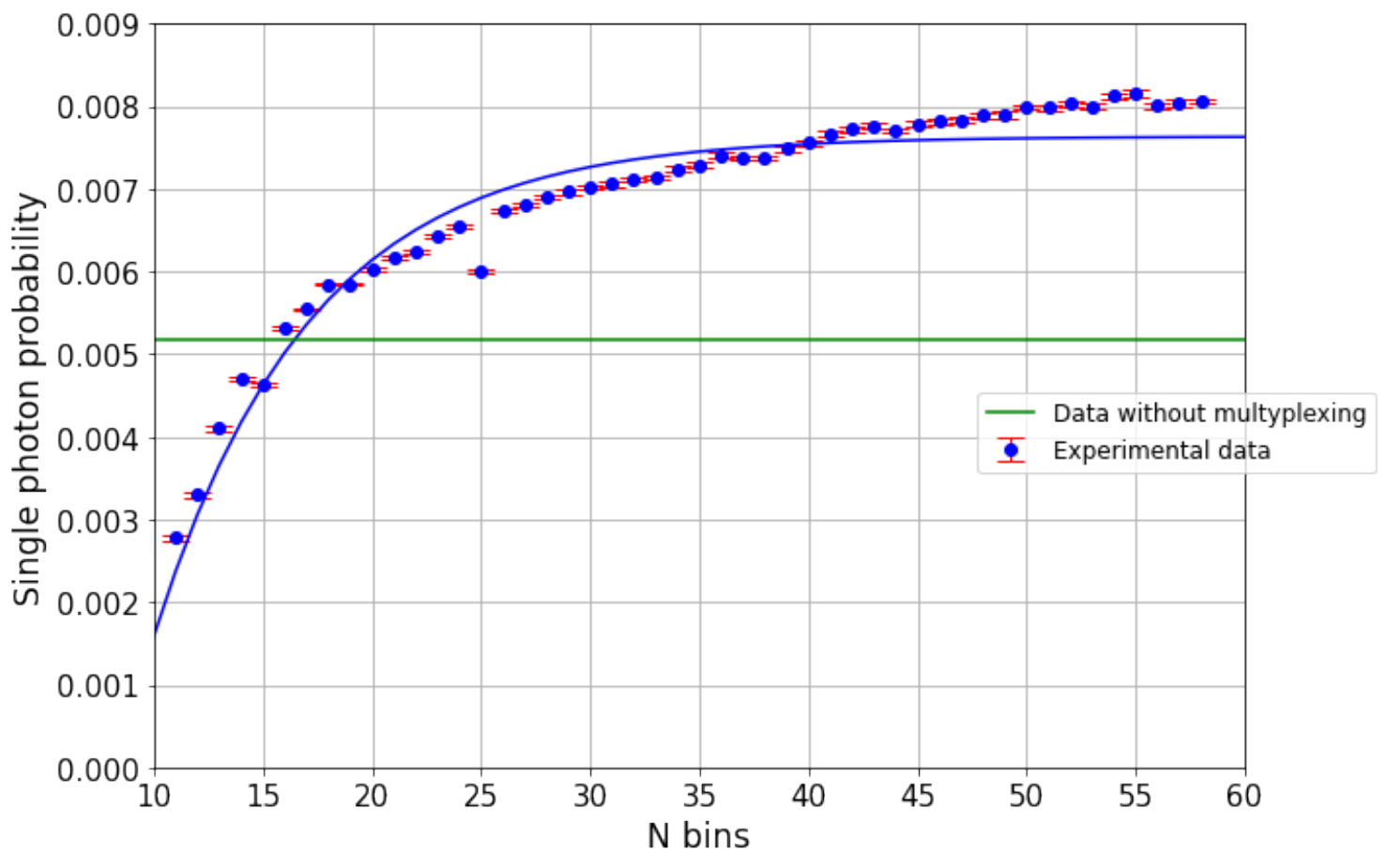}
  \caption{Single photon probability versus number of multiplexing modes. Blue dots match experimental obtained data with time multiplexing. Green line - experimental data without multiplexing. Blue line - approximation on experimental data by equal \eqref{eq:P_multiplexed} \label{fig:single photon prob}}
\end{figure}

As demonstrated in the Figure \ref{fig:single photon prob}, the probability of single-photon generation via time multiplexing surpasses that of the absence of multiplexing after 16 multiplexing modes. The blue line is an approximation of the experimental data, and it provides the following characteristics of the source: the average number of pairs per pump pulse $\mu \approx 6*10^{-3}$, heralding efficiency $\eta_s \approx 0.31$, losses per loop $\eta_{sl} \approx 0.067$.

\section{Conclusion}
A detailed analysis and experimental results presented in this article demonstrate that the development of a time-multiplexed heralded single-photon source using spontaneous parametric down-conversion provides a promising route to achieving near-deterministic single-photon generation. The employment of time-multiplexing techniques enhances the probability of photon generation. 

Theoretical and numerical models, when considered in conjunction with experimental observations, demonstrate the critical role of key parameters such as purity and heralding efficiency in determining the quality of the photon source. Furthermore, it has been demonstrated that the focal parameters of the SPDC crystal play an essential role in the optimisation of the heralding efficiency, but that compromise between photon generation efficiency and the photon generation efficiency is necessary.

The experimental configuration demonstrates advantageous heralding efficiency and low optical losses. The dependence of the probability of photon detection on the number of multiplexed modes indicates that further refinements in system parameters, such as reducing losses in optical memory cells and increasing generation efficiency, will lead to an increase in the generation probability.

In summary, the present work proposes a methodology for the generation of heralded single photons through time-multiplexing, with considerable ramifications for quantum communication, quantum computing, and high-precision quantum sensing. Further optimisation of the system and integration with existing photonic technologies could pave the way for scalable, high-efficiency single-photon sources that are essential for a broad range of quantum technologies.

\bibliography{main}

\begin{thebibliography}{37}%
\makeatletter
\providecommand \@ifxundefined [1]{%
 \@ifx{#1\undefined}
}%
\providecommand \@ifnum [1]{%
 \ifnum #1\expandafter \@firstoftwo
 \else \expandafter \@secondoftwo
 \fi
}%
\providecommand \@ifx [1]{%
 \ifx #1\expandafter \@firstoftwo
 \else \expandafter \@secondoftwo
 \fi
}%
\providecommand \natexlab [1]{#1}%
\providecommand \enquote  [1]{``#1''}%
\providecommand \bibnamefont  [1]{#1}%
\providecommand \bibfnamefont [1]{#1}%
\providecommand \citenamefont [1]{#1}%
\providecommand \href@noop [0]{\@secondoftwo}%
\providecommand \href [0]{\begingroup \@sanitize@url \@href}%
\providecommand \@href[1]{\@@startlink{#1}\@@href}%
\providecommand \@@href[1]{\endgroup#1\@@endlink}%
\providecommand \@sanitize@url [0]{\catcode `\\12\catcode `\$12\catcode `\&12\catcode `\#12\catcode `\^12\catcode `\_12\catcode `\%12\relax}%
\providecommand \@@startlink[1]{}%
\providecommand \@@endlink[0]{}%
\providecommand \url  [0]{\begingroup\@sanitize@url \@url }%
\providecommand \@url [1]{\endgroup\@href {#1}{\urlprefix }}%
\providecommand \urlprefix  [0]{URL }%
\providecommand \Eprint [0]{\href }%
\providecommand \doibase [0]{http://dx.doi.org/}%
\providecommand \selectlanguage [0]{\@gobble}%
\providecommand \bibinfo  [0]{\@secondoftwo}%
\providecommand \bibfield  [0]{\@secondoftwo}%
\providecommand \translation [1]{[#1]}%
\providecommand \BibitemOpen [0]{}%
\providecommand \bibitemStop [0]{}%
\providecommand \bibitemNoStop [0]{.\EOS\space}%
\providecommand \EOS [0]{\spacefactor3000\relax}%
\providecommand \BibitemShut  [1]{\csname bibitem#1\endcsname}%
\let\auto@bib@innerbib\@empty
\bibitem [{\citenamefont {Pittman}, \citenamefont {Jacobs},\ and\ \citenamefont {Franson}(2002)}]{pittman2002single}%
  \BibitemOpen
  \bibfield  {author} {\bibinfo {author} {\bibfnamefont {T.}~\bibnamefont {Pittman}}, \bibinfo {author} {\bibfnamefont {B.}~\bibnamefont {Jacobs}}, \ and\ \bibinfo {author} {\bibfnamefont {J.}~\bibnamefont {Franson}},\ }\bibfield  {title} {\enquote {\bibinfo {title} {Single photons on pseudodemand from stored parametric down-conversion},}\ }\href@noop {} {\bibfield  {journal} {\bibinfo  {journal} {Physical Review A}\ }\textbf {\bibinfo {volume} {66}},\ \bibinfo {pages} {042303} (\bibinfo {year} {2002})}\BibitemShut {NoStop}%
\bibitem [{\citenamefont {Monroe}(2002)}]{monroe2002quantum}%
  \BibitemOpen
  \bibfield  {author} {\bibinfo {author} {\bibfnamefont {C.}~\bibnamefont {Monroe}},\ }\bibfield  {title} {\enquote {\bibinfo {title} {Quantum information processing with atoms and photons},}\ }\href@noop {} {\bibfield  {journal} {\bibinfo  {journal} {Nature}\ }\textbf {\bibinfo {volume} {416}},\ \bibinfo {pages} {238--246} (\bibinfo {year} {2002})}\BibitemShut {NoStop}%
\bibitem [{\citenamefont {Northup}\ and\ \citenamefont {Blatt}(2014)}]{northup2014quantum}%
  \BibitemOpen
  \bibfield  {author} {\bibinfo {author} {\bibfnamefont {T.}~\bibnamefont {Northup}}\ and\ \bibinfo {author} {\bibfnamefont {R.}~\bibnamefont {Blatt}},\ }\bibfield  {title} {\enquote {\bibinfo {title} {Quantum information transfer using photons},}\ }\href@noop {} {\bibfield  {journal} {\bibinfo  {journal} {Nature photonics}\ }\textbf {\bibinfo {volume} {8}},\ \bibinfo {pages} {356--363} (\bibinfo {year} {2014})}\BibitemShut {NoStop}%
\bibitem [{\citenamefont {Afzelius}, \citenamefont {Gisin},\ and\ \citenamefont {De~Riedmatten}(2015)}]{afzelius2015quantum}%
  \BibitemOpen
  \bibfield  {author} {\bibinfo {author} {\bibfnamefont {M.}~\bibnamefont {Afzelius}}, \bibinfo {author} {\bibfnamefont {N.}~\bibnamefont {Gisin}}, \ and\ \bibinfo {author} {\bibfnamefont {H.}~\bibnamefont {De~Riedmatten}},\ }\bibfield  {title} {\enquote {\bibinfo {title} {Quantum memory for photons},}\ }\href@noop {} {\bibfield  {journal} {\bibinfo  {journal} {Physics Today}\ }\textbf {\bibinfo {volume} {68}},\ \bibinfo {pages} {42--47} (\bibinfo {year} {2015})}\BibitemShut {NoStop}%
\bibitem [{\citenamefont {Wang}\ \emph {et~al.}(2019)\citenamefont {Wang}, \citenamefont {Li}, \citenamefont {Zhang}, \citenamefont {Su}, \citenamefont {Zhou}, \citenamefont {Liao}, \citenamefont {Du}, \citenamefont {Yan},\ and\ \citenamefont {Zhu}}]{wang2019efficient}%
  \BibitemOpen
  \bibfield  {author} {\bibinfo {author} {\bibfnamefont {Y.}~\bibnamefont {Wang}}, \bibinfo {author} {\bibfnamefont {J.}~\bibnamefont {Li}}, \bibinfo {author} {\bibfnamefont {S.}~\bibnamefont {Zhang}}, \bibinfo {author} {\bibfnamefont {K.}~\bibnamefont {Su}}, \bibinfo {author} {\bibfnamefont {Y.}~\bibnamefont {Zhou}}, \bibinfo {author} {\bibfnamefont {K.}~\bibnamefont {Liao}}, \bibinfo {author} {\bibfnamefont {S.}~\bibnamefont {Du}}, \bibinfo {author} {\bibfnamefont {H.}~\bibnamefont {Yan}}, \ and\ \bibinfo {author} {\bibfnamefont {S.-L.}\ \bibnamefont {Zhu}},\ }\bibfield  {title} {\enquote {\bibinfo {title} {Efficient quantum memory for single-photon polarization qubits},}\ }\href@noop {} {\bibfield  {journal} {\bibinfo  {journal} {Nature Photonics}\ }\textbf {\bibinfo {volume} {13}},\ \bibinfo {pages} {346--351} (\bibinfo {year} {2019})}\BibitemShut {NoStop}%
\bibitem [{\citenamefont {Pirandola}\ \emph {et~al.}(2018)\citenamefont {Pirandola}, \citenamefont {Bardhan}, \citenamefont {Gehring}, \citenamefont {Weedbrook},\ and\ \citenamefont {Lloyd}}]{pirandola2018advances}%
  \BibitemOpen
  \bibfield  {author} {\bibinfo {author} {\bibfnamefont {S.}~\bibnamefont {Pirandola}}, \bibinfo {author} {\bibfnamefont {B.~R.}\ \bibnamefont {Bardhan}}, \bibinfo {author} {\bibfnamefont {T.}~\bibnamefont {Gehring}}, \bibinfo {author} {\bibfnamefont {C.}~\bibnamefont {Weedbrook}}, \ and\ \bibinfo {author} {\bibfnamefont {S.}~\bibnamefont {Lloyd}},\ }\bibfield  {title} {\enquote {\bibinfo {title} {Advances in photonic quantum sensing},}\ }\href@noop {} {\bibfield  {journal} {\bibinfo  {journal} {Nature Photonics}\ }\textbf {\bibinfo {volume} {12}},\ \bibinfo {pages} {724--733} (\bibinfo {year} {2018})}\BibitemShut {NoStop}%
\bibitem [{\citenamefont {Lawrie}\ \emph {et~al.}(2019)\citenamefont {Lawrie}, \citenamefont {Lett}, \citenamefont {Marino},\ and\ \citenamefont {Pooser}}]{lawrie2019quantum}%
  \BibitemOpen
  \bibfield  {author} {\bibinfo {author} {\bibfnamefont {B.~J.}\ \bibnamefont {Lawrie}}, \bibinfo {author} {\bibfnamefont {P.~D.}\ \bibnamefont {Lett}}, \bibinfo {author} {\bibfnamefont {A.~M.}\ \bibnamefont {Marino}}, \ and\ \bibinfo {author} {\bibfnamefont {R.~C.}\ \bibnamefont {Pooser}},\ }\bibfield  {title} {\enquote {\bibinfo {title} {Quantum sensing with squeezed light},}\ }\href@noop {} {\bibfield  {journal} {\bibinfo  {journal} {Acs Photonics}\ }\textbf {\bibinfo {volume} {6}},\ \bibinfo {pages} {1307--1318} (\bibinfo {year} {2019})}\BibitemShut {NoStop}%
\bibitem [{\citenamefont {Degen}, \citenamefont {Reinhard},\ and\ \citenamefont {Cappellaro}(2017)}]{degen2017quantum}%
  \BibitemOpen
  \bibfield  {author} {\bibinfo {author} {\bibfnamefont {C.~L.}\ \bibnamefont {Degen}}, \bibinfo {author} {\bibfnamefont {F.}~\bibnamefont {Reinhard}}, \ and\ \bibinfo {author} {\bibfnamefont {P.}~\bibnamefont {Cappellaro}},\ }\bibfield  {title} {\enquote {\bibinfo {title} {Quantum sensing},}\ }\href@noop {} {\bibfield  {journal} {\bibinfo  {journal} {Reviews of modern physics}\ }\textbf {\bibinfo {volume} {89}},\ \bibinfo {pages} {035002} (\bibinfo {year} {2017})}\BibitemShut {NoStop}%
\bibitem [{\citenamefont {Flamini}, \citenamefont {Spagnolo},\ and\ \citenamefont {Sciarrino}(2018)}]{flamini2018photonic}%
  \BibitemOpen
  \bibfield  {author} {\bibinfo {author} {\bibfnamefont {F.}~\bibnamefont {Flamini}}, \bibinfo {author} {\bibfnamefont {N.}~\bibnamefont {Spagnolo}}, \ and\ \bibinfo {author} {\bibfnamefont {F.}~\bibnamefont {Sciarrino}},\ }\bibfield  {title} {\enquote {\bibinfo {title} {Photonic quantum information processing: a review},}\ }\href@noop {} {\bibfield  {journal} {\bibinfo  {journal} {Reports on Progress in Physics}\ }\textbf {\bibinfo {volume} {82}},\ \bibinfo {pages} {016001} (\bibinfo {year} {2018})}\BibitemShut {NoStop}%
\bibitem [{\citenamefont {Hennrich}\ \emph {et~al.}(2004)\citenamefont {Hennrich}, \citenamefont {Legero}, \citenamefont {Kuhn},\ and\ \citenamefont {Rempe}}]{hennrich2004photon}%
  \BibitemOpen
  \bibfield  {author} {\bibinfo {author} {\bibfnamefont {M.}~\bibnamefont {Hennrich}}, \bibinfo {author} {\bibfnamefont {T.}~\bibnamefont {Legero}}, \bibinfo {author} {\bibfnamefont {A.}~\bibnamefont {Kuhn}}, \ and\ \bibinfo {author} {\bibfnamefont {G.}~\bibnamefont {Rempe}},\ }\bibfield  {title} {\enquote {\bibinfo {title} {Photon statistics of a non-stationary periodically driven single-photon source},}\ }\href@noop {} {\bibfield  {journal} {\bibinfo  {journal} {New Journal of Physics}\ }\textbf {\bibinfo {volume} {6}},\ \bibinfo {pages} {86} (\bibinfo {year} {2004})}\BibitemShut {NoStop}%
\bibitem [{\citenamefont {Hijlkema}\ \emph {et~al.}(2007)\citenamefont {Hijlkema}, \citenamefont {Weber}, \citenamefont {Specht}, \citenamefont {Webster}, \citenamefont {Kuhn},\ and\ \citenamefont {Rempe}}]{hijlkema2007single}%
  \BibitemOpen
  \bibfield  {author} {\bibinfo {author} {\bibfnamefont {M.}~\bibnamefont {Hijlkema}}, \bibinfo {author} {\bibfnamefont {B.}~\bibnamefont {Weber}}, \bibinfo {author} {\bibfnamefont {H.~P.}\ \bibnamefont {Specht}}, \bibinfo {author} {\bibfnamefont {S.~C.}\ \bibnamefont {Webster}}, \bibinfo {author} {\bibfnamefont {A.}~\bibnamefont {Kuhn}}, \ and\ \bibinfo {author} {\bibfnamefont {G.}~\bibnamefont {Rempe}},\ }\bibfield  {title} {\enquote {\bibinfo {title} {A single-photon server with just one atom},}\ }\href@noop {} {\bibfield  {journal} {\bibinfo  {journal} {Nature Physics}\ }\textbf {\bibinfo {volume} {3}},\ \bibinfo {pages} {253--255} (\bibinfo {year} {2007})}\BibitemShut {NoStop}%
\bibitem [{\citenamefont {Keller}\ \emph {et~al.}(2004)\citenamefont {Keller}, \citenamefont {Lange}, \citenamefont {Hayasaka}, \citenamefont {Lange},\ and\ \citenamefont {Walther}}]{keller2004continuous}%
  \BibitemOpen
  \bibfield  {author} {\bibinfo {author} {\bibfnamefont {M.}~\bibnamefont {Keller}}, \bibinfo {author} {\bibfnamefont {B.}~\bibnamefont {Lange}}, \bibinfo {author} {\bibfnamefont {K.}~\bibnamefont {Hayasaka}}, \bibinfo {author} {\bibfnamefont {W.}~\bibnamefont {Lange}}, \ and\ \bibinfo {author} {\bibfnamefont {H.}~\bibnamefont {Walther}},\ }\bibfield  {title} {\enquote {\bibinfo {title} {Continuous generation of single photons with controlled waveform in an ion-trap cavity system},}\ }\href@noop {} {\bibfield  {journal} {\bibinfo  {journal} {Nature}\ }\textbf {\bibinfo {volume} {431}},\ \bibinfo {pages} {1075--1078} (\bibinfo {year} {2004})}\BibitemShut {NoStop}%
\bibitem [{\citenamefont {Barros}\ \emph {et~al.}(2009)\citenamefont {Barros}, \citenamefont {Stute}, \citenamefont {Northup}, \citenamefont {Russo}, \citenamefont {Schmidt},\ and\ \citenamefont {Blatt}}]{barros2009deterministic}%
  \BibitemOpen
  \bibfield  {author} {\bibinfo {author} {\bibfnamefont {H.}~\bibnamefont {Barros}}, \bibinfo {author} {\bibfnamefont {A.}~\bibnamefont {Stute}}, \bibinfo {author} {\bibfnamefont {T.}~\bibnamefont {Northup}}, \bibinfo {author} {\bibfnamefont {C.}~\bibnamefont {Russo}}, \bibinfo {author} {\bibfnamefont {P.}~\bibnamefont {Schmidt}}, \ and\ \bibinfo {author} {\bibfnamefont {R.}~\bibnamefont {Blatt}},\ }\bibfield  {title} {\enquote {\bibinfo {title} {Deterministic single-photon source from a single ion},}\ }\href@noop {} {\bibfield  {journal} {\bibinfo  {journal} {New Journal of Physics}\ }\textbf {\bibinfo {volume} {11}},\ \bibinfo {pages} {103004} (\bibinfo {year} {2009})}\BibitemShut {NoStop}%
\bibitem [{\citenamefont {Maurer}\ \emph {et~al.}(2004)\citenamefont {Maurer}, \citenamefont {Becher}, \citenamefont {Russo}, \citenamefont {Eschner},\ and\ \citenamefont {Blatt}}]{maurer2004single}%
  \BibitemOpen
  \bibfield  {author} {\bibinfo {author} {\bibfnamefont {C.}~\bibnamefont {Maurer}}, \bibinfo {author} {\bibfnamefont {C.}~\bibnamefont {Becher}}, \bibinfo {author} {\bibfnamefont {C.}~\bibnamefont {Russo}}, \bibinfo {author} {\bibfnamefont {J.}~\bibnamefont {Eschner}}, \ and\ \bibinfo {author} {\bibfnamefont {R.}~\bibnamefont {Blatt}},\ }\bibfield  {title} {\enquote {\bibinfo {title} {A single-photon source based on a single ca+ ion},}\ }\href@noop {} {\bibfield  {journal} {\bibinfo  {journal} {New journal of physics}\ }\textbf {\bibinfo {volume} {6}},\ \bibinfo {pages} {94} (\bibinfo {year} {2004})}\BibitemShut {NoStop}%
\bibitem [{\citenamefont {Kiraz}\ \emph {et~al.}(2005)\citenamefont {Kiraz}, \citenamefont {Ehrl}, \citenamefont {Hellerer}, \citenamefont {M{\"u}stecapl{\i}o{\u{g}}lu}, \citenamefont {Br{\"a}uchle},\ and\ \citenamefont {Zumbusch}}]{kiraz2005indistinguishable}%
  \BibitemOpen
  \bibfield  {author} {\bibinfo {author} {\bibfnamefont {A.}~\bibnamefont {Kiraz}}, \bibinfo {author} {\bibfnamefont {M.}~\bibnamefont {Ehrl}}, \bibinfo {author} {\bibfnamefont {T.}~\bibnamefont {Hellerer}}, \bibinfo {author} {\bibfnamefont {{\"O}.}~\bibnamefont {M{\"u}stecapl{\i}o{\u{g}}lu}}, \bibinfo {author} {\bibfnamefont {C.}~\bibnamefont {Br{\"a}uchle}}, \ and\ \bibinfo {author} {\bibfnamefont {A.}~\bibnamefont {Zumbusch}},\ }\bibfield  {title} {\enquote {\bibinfo {title} {Indistinguishable photons from a single molecule},}\ }\href@noop {} {\bibfield  {journal} {\bibinfo  {journal} {Physical review letters}\ }\textbf {\bibinfo {volume} {94}},\ \bibinfo {pages} {223602} (\bibinfo {year} {2005})}\BibitemShut {NoStop}%
\bibitem [{\citenamefont {Fleury}\ \emph {et~al.}(2000)\citenamefont {Fleury}, \citenamefont {Segura}, \citenamefont {Zumofen}, \citenamefont {Hecht},\ and\ \citenamefont {Wild}}]{fleury2000nonclassical}%
  \BibitemOpen
  \bibfield  {author} {\bibinfo {author} {\bibfnamefont {L.}~\bibnamefont {Fleury}}, \bibinfo {author} {\bibfnamefont {J.-M.}\ \bibnamefont {Segura}}, \bibinfo {author} {\bibfnamefont {G.}~\bibnamefont {Zumofen}}, \bibinfo {author} {\bibfnamefont {B.}~\bibnamefont {Hecht}}, \ and\ \bibinfo {author} {\bibfnamefont {U.}~\bibnamefont {Wild}},\ }\bibfield  {title} {\enquote {\bibinfo {title} {Nonclassical photon statistics in single-molecule fluorescence at room temperature},}\ }\href@noop {} {\bibfield  {journal} {\bibinfo  {journal} {Physical review letters}\ }\textbf {\bibinfo {volume} {84}},\ \bibinfo {pages} {1148} (\bibinfo {year} {2000})}\BibitemShut {NoStop}%
\bibitem [{\citenamefont {Senellart}, \citenamefont {Solomon},\ and\ \citenamefont {White}(2017)}]{senellart2017high}%
  \BibitemOpen
  \bibfield  {author} {\bibinfo {author} {\bibfnamefont {P.}~\bibnamefont {Senellart}}, \bibinfo {author} {\bibfnamefont {G.}~\bibnamefont {Solomon}}, \ and\ \bibinfo {author} {\bibfnamefont {A.}~\bibnamefont {White}},\ }\bibfield  {title} {\enquote {\bibinfo {title} {High-performance semiconductor quantum-dot single-photon sources},}\ }\href@noop {} {\bibfield  {journal} {\bibinfo  {journal} {Nature nanotechnology}\ }\textbf {\bibinfo {volume} {12}},\ \bibinfo {pages} {1026--1039} (\bibinfo {year} {2017})}\BibitemShut {NoStop}%
\bibitem [{\citenamefont {Arakawa}\ and\ \citenamefont {Holmes}(2020)}]{arakawa2020progress}%
  \BibitemOpen
  \bibfield  {author} {\bibinfo {author} {\bibfnamefont {Y.}~\bibnamefont {Arakawa}}\ and\ \bibinfo {author} {\bibfnamefont {M.~J.}\ \bibnamefont {Holmes}},\ }\bibfield  {title} {\enquote {\bibinfo {title} {Progress in quantum-dot single photon sources for quantum information technologies: A broad spectrum overview},}\ }\href@noop {} {\bibfield  {journal} {\bibinfo  {journal} {Applied Physics Reviews}\ }\textbf {\bibinfo {volume} {7}} (\bibinfo {year} {2020})}\BibitemShut {NoStop}%
\bibitem [{\citenamefont {Aharonovich}\ \emph {et~al.}(2011)\citenamefont {Aharonovich}, \citenamefont {Castelletto}, \citenamefont {Simpson}, \citenamefont {Su}, \citenamefont {Greentree},\ and\ \citenamefont {Prawer}}]{aharonovich2011diamond}%
  \BibitemOpen
  \bibfield  {author} {\bibinfo {author} {\bibfnamefont {I.}~\bibnamefont {Aharonovich}}, \bibinfo {author} {\bibfnamefont {S.}~\bibnamefont {Castelletto}}, \bibinfo {author} {\bibfnamefont {D.}~\bibnamefont {Simpson}}, \bibinfo {author} {\bibfnamefont {C.-H.}\ \bibnamefont {Su}}, \bibinfo {author} {\bibfnamefont {A.}~\bibnamefont {Greentree}}, \ and\ \bibinfo {author} {\bibfnamefont {S.}~\bibnamefont {Prawer}},\ }\bibfield  {title} {\enquote {\bibinfo {title} {Diamond-based single-photon emitters},}\ }\href@noop {} {\bibfield  {journal} {\bibinfo  {journal} {Reports on progress in Physics}\ }\textbf {\bibinfo {volume} {74}},\ \bibinfo {pages} {076501} (\bibinfo {year} {2011})}\BibitemShut {NoStop}%
\bibitem [{\citenamefont {Aharonovich}, \citenamefont {Englund},\ and\ \citenamefont {Toth}(2016)}]{aharonovich2016solid}%
  \BibitemOpen
  \bibfield  {author} {\bibinfo {author} {\bibfnamefont {I.}~\bibnamefont {Aharonovich}}, \bibinfo {author} {\bibfnamefont {D.}~\bibnamefont {Englund}}, \ and\ \bibinfo {author} {\bibfnamefont {M.}~\bibnamefont {Toth}},\ }\bibfield  {title} {\enquote {\bibinfo {title} {Solid-state single-photon emitters},}\ }\href@noop {} {\bibfield  {journal} {\bibinfo  {journal} {Nature photonics}\ }\textbf {\bibinfo {volume} {10}},\ \bibinfo {pages} {631--641} (\bibinfo {year} {2016})}\BibitemShut {NoStop}%
\bibitem [{\citenamefont {Kaneda}\ \emph {et~al.}(2015)\citenamefont {Kaneda}, \citenamefont {Christensen}, \citenamefont {Wong}, \citenamefont {Park}, \citenamefont {McCusker},\ and\ \citenamefont {Kwiat}}]{kaneda2015time}%
  \BibitemOpen
  \bibfield  {author} {\bibinfo {author} {\bibfnamefont {F.}~\bibnamefont {Kaneda}}, \bibinfo {author} {\bibfnamefont {B.~G.}\ \bibnamefont {Christensen}}, \bibinfo {author} {\bibfnamefont {J.~J.}\ \bibnamefont {Wong}}, \bibinfo {author} {\bibfnamefont {H.~S.}\ \bibnamefont {Park}}, \bibinfo {author} {\bibfnamefont {K.~T.}\ \bibnamefont {McCusker}}, \ and\ \bibinfo {author} {\bibfnamefont {P.~G.}\ \bibnamefont {Kwiat}},\ }\bibfield  {title} {\enquote {\bibinfo {title} {Time-multiplexed heralded single-photon source},}\ }\href@noop {} {\bibfield  {journal} {\bibinfo  {journal} {Optica}\ }\textbf {\bibinfo {volume} {2}},\ \bibinfo {pages} {1010--1013} (\bibinfo {year} {2015})}\BibitemShut {NoStop}%
\bibitem [{\citenamefont {Ma}\ \emph {et~al.}(2011)\citenamefont {Ma}, \citenamefont {Zotter}, \citenamefont {Kofler}, \citenamefont {Jennewein},\ and\ \citenamefont {Zeilinger}}]{ma2011experimental}%
  \BibitemOpen
  \bibfield  {author} {\bibinfo {author} {\bibfnamefont {X.-s.}\ \bibnamefont {Ma}}, \bibinfo {author} {\bibfnamefont {S.}~\bibnamefont {Zotter}}, \bibinfo {author} {\bibfnamefont {J.}~\bibnamefont {Kofler}}, \bibinfo {author} {\bibfnamefont {T.}~\bibnamefont {Jennewein}}, \ and\ \bibinfo {author} {\bibfnamefont {A.}~\bibnamefont {Zeilinger}},\ }\bibfield  {title} {\enquote {\bibinfo {title} {Experimental generation of single photons via active multiplexing},}\ }\href@noop {} {\bibfield  {journal} {\bibinfo  {journal} {Physical Review A—Atomic, Molecular, and Optical Physics}\ }\textbf {\bibinfo {volume} {83}},\ \bibinfo {pages} {043814} (\bibinfo {year} {2011})}\BibitemShut {NoStop}%
\bibitem [{\citenamefont {Magnitskiy}\ \emph {et~al.}(2015)\citenamefont {Magnitskiy}, \citenamefont {Frolovtsev}, \citenamefont {Firsov}, \citenamefont {Gostev}, \citenamefont {Protsenko},\ and\ \citenamefont {Saygin}}]{magnitskiy2015spdc}%
  \BibitemOpen
  \bibfield  {author} {\bibinfo {author} {\bibfnamefont {S.}~\bibnamefont {Magnitskiy}}, \bibinfo {author} {\bibfnamefont {D.}~\bibnamefont {Frolovtsev}}, \bibinfo {author} {\bibfnamefont {V.}~\bibnamefont {Firsov}}, \bibinfo {author} {\bibfnamefont {P.}~\bibnamefont {Gostev}}, \bibinfo {author} {\bibfnamefont {I.}~\bibnamefont {Protsenko}}, \ and\ \bibinfo {author} {\bibfnamefont {M.}~\bibnamefont {Saygin}},\ }\bibfield  {title} {\enquote {\bibinfo {title} {A spdc-based source of entangled photons and its characterization},}\ }\href@noop {} {\bibfield  {journal} {\bibinfo  {journal} {Journal of Russian Laser Research}\ }\textbf {\bibinfo {volume} {36}},\ \bibinfo {pages} {618--629} (\bibinfo {year} {2015})}\BibitemShut {NoStop}%
\bibitem [{\citenamefont {Goldschmidt}\ \emph {et~al.}(2008)\citenamefont {Goldschmidt}, \citenamefont {Eisaman}, \citenamefont {Fan}, \citenamefont {Polyakov},\ and\ \citenamefont {Migdall}}]{goldschmidt2008spectrally}%
  \BibitemOpen
  \bibfield  {author} {\bibinfo {author} {\bibfnamefont {E.~A.}\ \bibnamefont {Goldschmidt}}, \bibinfo {author} {\bibfnamefont {M.~D.}\ \bibnamefont {Eisaman}}, \bibinfo {author} {\bibfnamefont {J.}~\bibnamefont {Fan}}, \bibinfo {author} {\bibfnamefont {S.~V.}\ \bibnamefont {Polyakov}}, \ and\ \bibinfo {author} {\bibfnamefont {A.}~\bibnamefont {Migdall}},\ }\bibfield  {title} {\enquote {\bibinfo {title} {Spectrally bright and broad fiber-based heralded single-photon source},}\ }\href@noop {} {\bibfield  {journal} {\bibinfo  {journal} {Physical Review A—Atomic, Molecular, and Optical Physics}\ }\textbf {\bibinfo {volume} {78}},\ \bibinfo {pages} {013844} (\bibinfo {year} {2008})}\BibitemShut {NoStop}%
\bibitem [{\citenamefont {Fiorentino}\ \emph {et~al.}(2002)\citenamefont {Fiorentino}, \citenamefont {Voss}, \citenamefont {Sharping},\ and\ \citenamefont {Kumar}}]{fiorentino2002all}%
  \BibitemOpen
  \bibfield  {author} {\bibinfo {author} {\bibfnamefont {M.}~\bibnamefont {Fiorentino}}, \bibinfo {author} {\bibfnamefont {P.~L.}\ \bibnamefont {Voss}}, \bibinfo {author} {\bibfnamefont {J.~E.}\ \bibnamefont {Sharping}}, \ and\ \bibinfo {author} {\bibfnamefont {P.}~\bibnamefont {Kumar}},\ }\bibfield  {title} {\enquote {\bibinfo {title} {All-fiber photon-pair source for quantum communications},}\ }\href@noop {} {\bibfield  {journal} {\bibinfo  {journal} {IEEE Photonics Technology Letters}\ }\textbf {\bibinfo {volume} {14}},\ \bibinfo {pages} {983--985} (\bibinfo {year} {2002})}\BibitemShut {NoStop}%
\bibitem [{\citenamefont {Li}\ \emph {et~al.}(2005)\citenamefont {Li}, \citenamefont {Voss}, \citenamefont {Sharping},\ and\ \citenamefont {Kumar}}]{li2005optical}%
  \BibitemOpen
  \bibfield  {author} {\bibinfo {author} {\bibfnamefont {X.}~\bibnamefont {Li}}, \bibinfo {author} {\bibfnamefont {P.~L.}\ \bibnamefont {Voss}}, \bibinfo {author} {\bibfnamefont {J.~E.}\ \bibnamefont {Sharping}}, \ and\ \bibinfo {author} {\bibfnamefont {P.}~\bibnamefont {Kumar}},\ }\bibfield  {title} {\enquote {\bibinfo {title} {Optical-fiber source of polarization-entangled photons in the 1550 nm telecom band},}\ }\href@noop {} {\bibfield  {journal} {\bibinfo  {journal} {Physical review letters}\ }\textbf {\bibinfo {volume} {94}},\ \bibinfo {pages} {053601} (\bibinfo {year} {2005})}\BibitemShut {NoStop}%
\bibitem [{\citenamefont {Christ}\ and\ \citenamefont {Silberhorn}(2012)}]{christ2012limits}%
  \BibitemOpen
  \bibfield  {author} {\bibinfo {author} {\bibfnamefont {A.}~\bibnamefont {Christ}}\ and\ \bibinfo {author} {\bibfnamefont {C.}~\bibnamefont {Silberhorn}},\ }\bibfield  {title} {\enquote {\bibinfo {title} {Limits on the deterministic creation of pure single-photon states using parametric down-conversion},}\ }\href@noop {} {\bibfield  {journal} {\bibinfo  {journal} {Physical Review A—Atomic, Molecular, and Optical Physics}\ }\textbf {\bibinfo {volume} {85}},\ \bibinfo {pages} {023829} (\bibinfo {year} {2012})}\BibitemShut {NoStop}%
\bibitem [{\citenamefont {Joshi}\ \emph {et~al.}(2018)\citenamefont {Joshi}, \citenamefont {Farsi}, \citenamefont {Clemmen}, \citenamefont {Ramelow},\ and\ \citenamefont {Gaeta}}]{joshi2018frequency}%
  \BibitemOpen
  \bibfield  {author} {\bibinfo {author} {\bibfnamefont {C.}~\bibnamefont {Joshi}}, \bibinfo {author} {\bibfnamefont {A.}~\bibnamefont {Farsi}}, \bibinfo {author} {\bibfnamefont {S.}~\bibnamefont {Clemmen}}, \bibinfo {author} {\bibfnamefont {S.}~\bibnamefont {Ramelow}}, \ and\ \bibinfo {author} {\bibfnamefont {A.~L.}\ \bibnamefont {Gaeta}},\ }\bibfield  {title} {\enquote {\bibinfo {title} {Frequency multiplexing for quasi-deterministic heralded single-photon sources},}\ }\href@noop {} {\bibfield  {journal} {\bibinfo  {journal} {Nature communications}\ }\textbf {\bibinfo {volume} {9}},\ \bibinfo {pages} {847} (\bibinfo {year} {2018})}\BibitemShut {NoStop}%
\bibitem [{\citenamefont {Jeffrey}, \citenamefont {Peters},\ and\ \citenamefont {Kwiat}(2004)}]{jeffrey2004towards}%
  \BibitemOpen
  \bibfield  {author} {\bibinfo {author} {\bibfnamefont {E.}~\bibnamefont {Jeffrey}}, \bibinfo {author} {\bibfnamefont {N.~A.}\ \bibnamefont {Peters}}, \ and\ \bibinfo {author} {\bibfnamefont {P.~G.}\ \bibnamefont {Kwiat}},\ }\bibfield  {title} {\enquote {\bibinfo {title} {Towards a periodic deterministic source of arbitrary single-photon states},}\ }\href@noop {} {\bibfield  {journal} {\bibinfo  {journal} {New Journal of Physics}\ }\textbf {\bibinfo {volume} {6}},\ \bibinfo {pages} {100} (\bibinfo {year} {2004})}\BibitemShut {NoStop}%
\bibitem [{\citenamefont {Kaneda}\ and\ \citenamefont {Kwiat}(2019)}]{kaneda2019high}%
  \BibitemOpen
  \bibfield  {author} {\bibinfo {author} {\bibfnamefont {F.}~\bibnamefont {Kaneda}}\ and\ \bibinfo {author} {\bibfnamefont {P.~G.}\ \bibnamefont {Kwiat}},\ }\bibfield  {title} {\enquote {\bibinfo {title} {High-efficiency single-photon generation via large-scale active time multiplexing},}\ }\href@noop {} {\bibfield  {journal} {\bibinfo  {journal} {Science advances}\ }\textbf {\bibinfo {volume} {5}},\ \bibinfo {pages} {eaaw8586} (\bibinfo {year} {2019})}\BibitemShut {NoStop}%
\bibitem [{\citenamefont {Bennink}(2010)}]{bennink2010optimal}%
  \BibitemOpen
  \bibfield  {author} {\bibinfo {author} {\bibfnamefont {R.~S.}\ \bibnamefont {Bennink}},\ }\bibfield  {title} {\enquote {\bibinfo {title} {Optimal collinear gaussian beams for spontaneous parametric down-conversion},}\ }\href@noop {} {\bibfield  {journal} {\bibinfo  {journal} {Physical Review A—Atomic, Molecular, and Optical Physics}\ }\textbf {\bibinfo {volume} {81}},\ \bibinfo {pages} {053805} (\bibinfo {year} {2010})}\BibitemShut {NoStop}%
\bibitem [{\citenamefont {Kovlakov}\ \emph {et~al.}(2017)\citenamefont {Kovlakov}, \citenamefont {Bobrov}, \citenamefont {Straupe},\ and\ \citenamefont {Kulik}}]{kovlakov2017spatial}%
  \BibitemOpen
  \bibfield  {author} {\bibinfo {author} {\bibfnamefont {E.}~\bibnamefont {Kovlakov}}, \bibinfo {author} {\bibfnamefont {I.}~\bibnamefont {Bobrov}}, \bibinfo {author} {\bibfnamefont {S.}~\bibnamefont {Straupe}}, \ and\ \bibinfo {author} {\bibfnamefont {S.}~\bibnamefont {Kulik}},\ }\bibfield  {title} {\enquote {\bibinfo {title} {Spatial bell-state generation without transverse mode subspace postselection},}\ }\href@noop {} {\bibfield  {journal} {\bibinfo  {journal} {Physical review letters}\ }\textbf {\bibinfo {volume} {118}},\ \bibinfo {pages} {030503} (\bibinfo {year} {2017})}\BibitemShut {NoStop}%
\bibitem [{\citenamefont {Kaneda}\ \emph {et~al.}(2016)\citenamefont {Kaneda}, \citenamefont {Garay-Palmett}, \citenamefont {U’Ren},\ and\ \citenamefont {Kwiat}}]{kaneda2016heralded}%
  \BibitemOpen
  \bibfield  {author} {\bibinfo {author} {\bibfnamefont {F.}~\bibnamefont {Kaneda}}, \bibinfo {author} {\bibfnamefont {K.}~\bibnamefont {Garay-Palmett}}, \bibinfo {author} {\bibfnamefont {A.~B.}\ \bibnamefont {U’Ren}}, \ and\ \bibinfo {author} {\bibfnamefont {P.~G.}\ \bibnamefont {Kwiat}},\ }\bibfield  {title} {\enquote {\bibinfo {title} {Heralded single-photon source utilizing highly nondegenerate, spectrally factorable spontaneous parametric downconversion},}\ }\href@noop {} {\bibfield  {journal} {\bibinfo  {journal} {Optics express}\ }\textbf {\bibinfo {volume} {24}},\ \bibinfo {pages} {10733--10747} (\bibinfo {year} {2016})}\BibitemShut {NoStop}%
\bibitem [{\citenamefont {Zielnicki}\ \emph {et~al.}(2018)\citenamefont {Zielnicki}, \citenamefont {Garay-Palmett}, \citenamefont {Cruz-Delgado}, \citenamefont {Cruz-Ramirez}, \citenamefont {O’Boyle}, \citenamefont {Fang}, \citenamefont {Lorenz}, \citenamefont {U’Ren},\ and\ \citenamefont {Kwiat}}]{zielnicki2018joint}%
  \BibitemOpen
  \bibfield  {author} {\bibinfo {author} {\bibfnamefont {K.}~\bibnamefont {Zielnicki}}, \bibinfo {author} {\bibfnamefont {K.}~\bibnamefont {Garay-Palmett}}, \bibinfo {author} {\bibfnamefont {D.}~\bibnamefont {Cruz-Delgado}}, \bibinfo {author} {\bibfnamefont {H.}~\bibnamefont {Cruz-Ramirez}}, \bibinfo {author} {\bibfnamefont {M.~F.}\ \bibnamefont {O’Boyle}}, \bibinfo {author} {\bibfnamefont {B.}~\bibnamefont {Fang}}, \bibinfo {author} {\bibfnamefont {V.~O.}\ \bibnamefont {Lorenz}}, \bibinfo {author} {\bibfnamefont {A.~B.}\ \bibnamefont {U’Ren}}, \ and\ \bibinfo {author} {\bibfnamefont {P.~G.}\ \bibnamefont {Kwiat}},\ }\bibfield  {title} {\enquote {\bibinfo {title} {Joint spectral characterization of photon-pair sources},}\ }\href@noop {} {\bibfield  {journal} {\bibinfo  {journal} {Journal of Modern Optics}\ }\textbf {\bibinfo {volume} {65}},\ \bibinfo {pages} {1141--1160} (\bibinfo {year} {2018})}\BibitemShut {NoStop}%
\bibitem [{\citenamefont {Pickston}\ \emph {et~al.}(2021)\citenamefont {Pickston}, \citenamefont {Graffitti}, \citenamefont {Barrow}, \citenamefont {Morrison}, \citenamefont {Ho}, \citenamefont {Bra{\'n}czyk},\ and\ \citenamefont {Fedrizzi}}]{pickston2021optimised}%
  \BibitemOpen
  \bibfield  {author} {\bibinfo {author} {\bibfnamefont {A.}~\bibnamefont {Pickston}}, \bibinfo {author} {\bibfnamefont {F.}~\bibnamefont {Graffitti}}, \bibinfo {author} {\bibfnamefont {P.}~\bibnamefont {Barrow}}, \bibinfo {author} {\bibfnamefont {C.~L.}\ \bibnamefont {Morrison}}, \bibinfo {author} {\bibfnamefont {J.}~\bibnamefont {Ho}}, \bibinfo {author} {\bibfnamefont {A.~M.}\ \bibnamefont {Bra{\'n}czyk}}, \ and\ \bibinfo {author} {\bibfnamefont {A.}~\bibnamefont {Fedrizzi}},\ }\bibfield  {title} {\enquote {\bibinfo {title} {{Optimised domain-engineered crystals for pure telecom photon sources}},}\ }\href@noop {} {\bibfield  {journal} {\bibinfo  {journal} {Optics Express}\ }\textbf {\bibinfo {volume} {29}},\ \bibinfo {pages} {6991--7002} (\bibinfo {year} {2021})}\BibitemShut {NoStop}%
\bibitem [{\citenamefont {Tambasco}\ \emph {et~al.}(2016)\citenamefont {Tambasco}, \citenamefont {Boes}, \citenamefont {Helt}, \citenamefont {Steel},\ and\ \citenamefont {Mitchell}}]{tambasco2016domain}%
  \BibitemOpen
  \bibfield  {author} {\bibinfo {author} {\bibfnamefont {J.-L.}\ \bibnamefont {Tambasco}}, \bibinfo {author} {\bibfnamefont {A.}~\bibnamefont {Boes}}, \bibinfo {author} {\bibfnamefont {L.}~\bibnamefont {Helt}}, \bibinfo {author} {\bibfnamefont {M.}~\bibnamefont {Steel}}, \ and\ \bibinfo {author} {\bibfnamefont {A.}~\bibnamefont {Mitchell}},\ }\bibfield  {title} {\enquote {\bibinfo {title} {Domain engineering algorithm for practical and effective photon sources},}\ }\href@noop {} {\bibfield  {journal} {\bibinfo  {journal} {Optics express}\ }\textbf {\bibinfo {volume} {24}},\ \bibinfo {pages} {19616--19626} (\bibinfo {year} {2016})}\BibitemShut {NoStop}%
\bibitem [{\citenamefont {Graffitti}\ \emph {et~al.}(2017)\citenamefont {Graffitti}, \citenamefont {Kundys}, \citenamefont {Reid}, \citenamefont {Bra{\'n}czyk},\ and\ \citenamefont {Fedrizzi}}]{graffitti2017pure}%
  \BibitemOpen
  \bibfield  {author} {\bibinfo {author} {\bibfnamefont {F.}~\bibnamefont {Graffitti}}, \bibinfo {author} {\bibfnamefont {D.}~\bibnamefont {Kundys}}, \bibinfo {author} {\bibfnamefont {D.~T.}\ \bibnamefont {Reid}}, \bibinfo {author} {\bibfnamefont {A.~M.}\ \bibnamefont {Bra{\'n}czyk}}, \ and\ \bibinfo {author} {\bibfnamefont {A.}~\bibnamefont {Fedrizzi}},\ }\bibfield  {title} {\enquote {\bibinfo {title} {Pure down-conversion photons through sub-coherence-length domain engineering},}\ }\href@noop {} {\bibfield  {journal} {\bibinfo  {journal} {Quantum Science and Technology}\ }\textbf {\bibinfo {volume} {2}},\ \bibinfo {pages} {035001} (\bibinfo {year} {2017})}\BibitemShut {NoStop}%
\end{thebibliography}%

\section{Appendix A}

In this appendix we discusses about fairness of simplifications \eqref{eq:approx_1} and \eqref{eq:approx_2}:
\begin{eqnarray} \label{eq:approx_1_appendix}
    &&C \approx 0 \\
    &&A_+,B_+,\xi \approx const \label{eq:approx_2_appendix}
\end{eqnarray}

\begin{figure}[!h]
  \centering
  \includegraphics[width=0.99\linewidth]{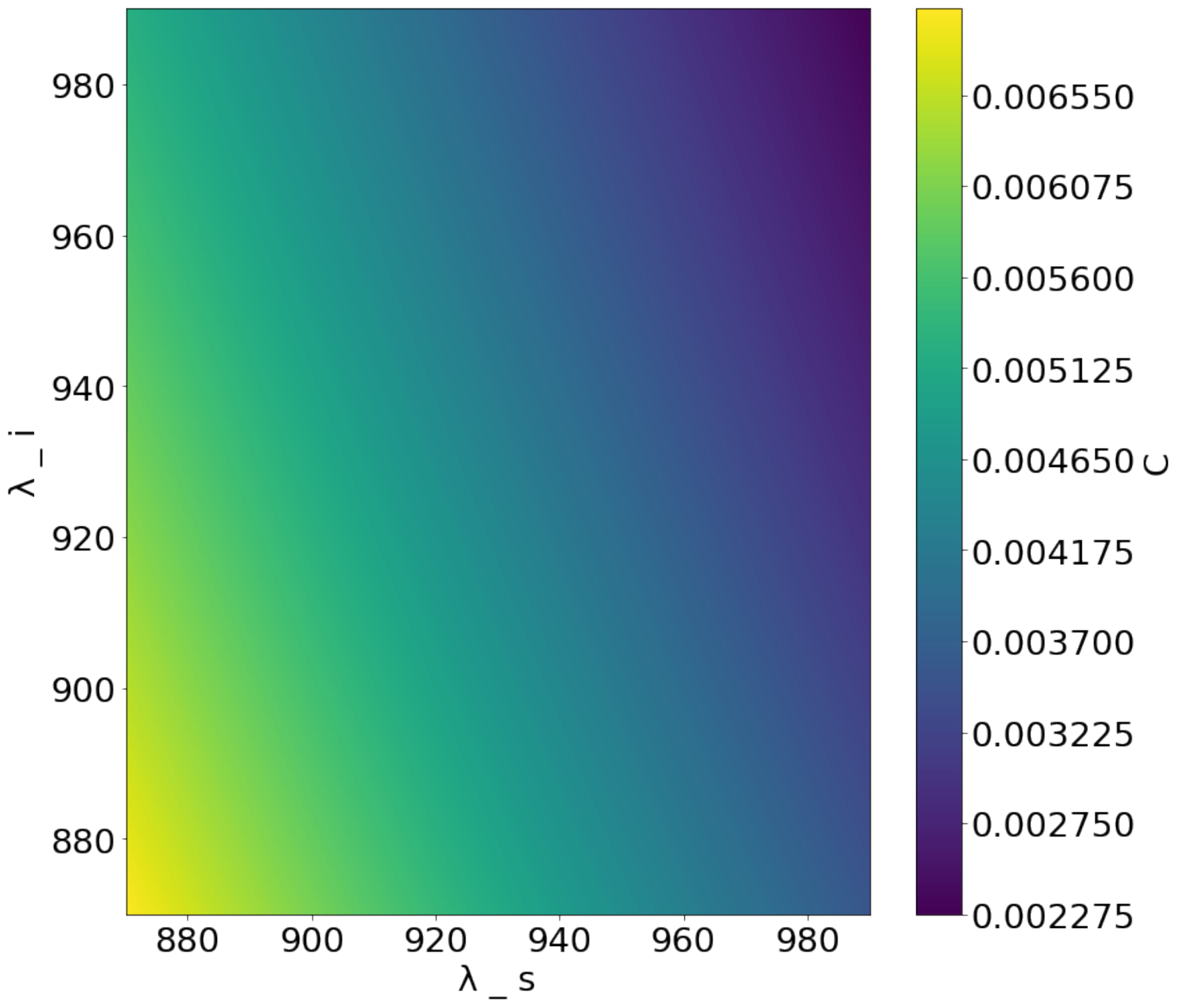},
  \caption{Approximation $C\approx0$\label{fig:C_plus}}
\end{figure}

\begin{figure}[!h]
  \centering
  \includegraphics[width=0.99\linewidth]{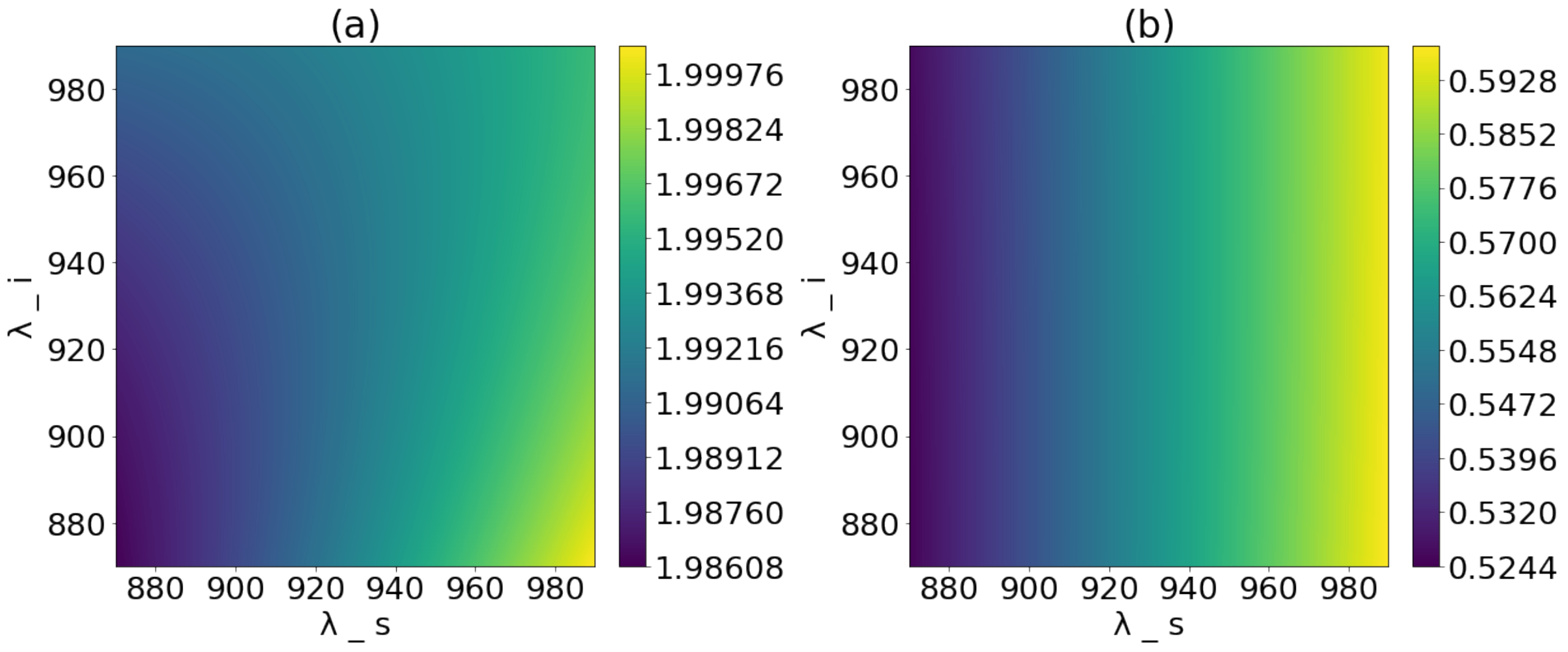},
  \caption{Wavelength dependence for (a) -- $B_+$, (b) -- $\xi$\label{fig:Approximation_2}}
\end{figure}

Equation \eqref{eq:approx_1_appendix} facilitates the simplification of integration in \eqref{eq:JSA}. When $|C|\lesssim 0.1$ near-phase matching conditions are approved. In our case $C\thicksim10^{-3}$, as illustrated in figure \ref{fig:C_plus}, thereby validating the approximation \eqref{eq:approx_1_appendix}.

Another approximation \eqref{eq:approx_2_appendix} is considered in the range ($\omega_s,\omega_i$) on which the amplitude of \eqref{eq:JSA} is appreciable. This approximation \eqref{eq:approx_2_appendix} indicates that the frequency dependence is predominantly dictated by the pump spectrum $s(\omega_p)$ and the phase mismatch parameter $\Delta k$. A closer examination of equations \eqref{eq:A_+} and \eqref{eq:focal_par} reveals that $A_+$ is entirely dependent on the size of the beam waist. Consequently, under constant beam waists, $A_+$ is a constant. The wavelength dependence of $B_+$ and $\xi$ is illustrated in Figure \ref{fig:Approximation_2}. The approximation in equation \eqref{eq:approx_2_appendix} can therefore be considered valid for our parameters.

\section{Appendix B}

\begin{figure*}
  \centering
  \includegraphics[width=0.99\linewidth]{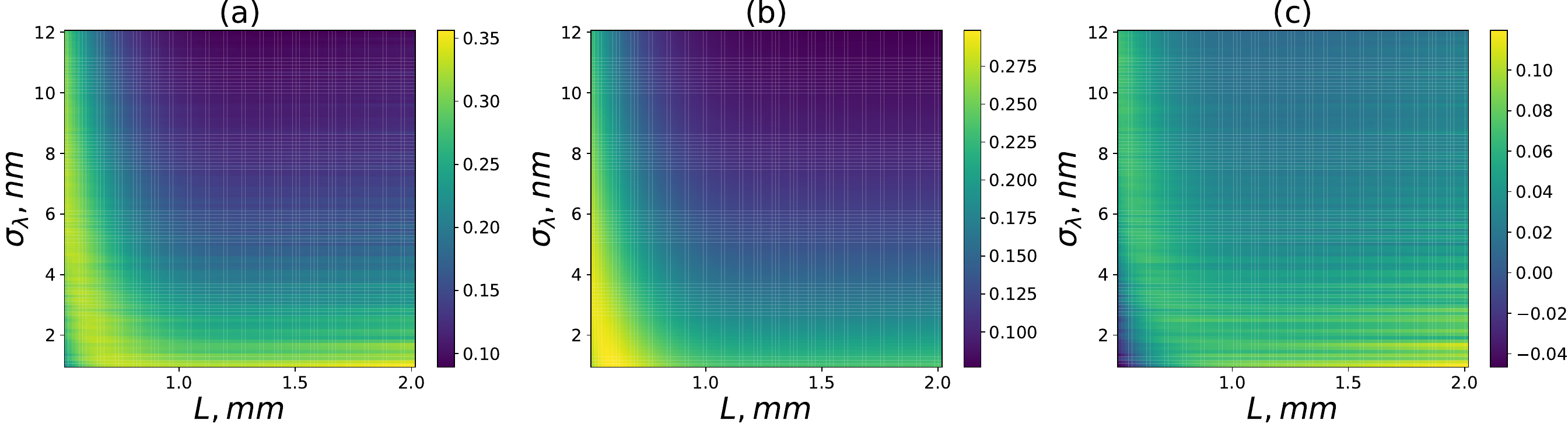},
  \caption{A plot of purity values is provided for various spectral pump widths and crystal lengths. \label{fig:Puriry from siglam length2}}
\end{figure*}

This section discusses optimizing the domain structure of crystals for high purity in general.
Consider the widespread description of JSA: $\psi(\omega_s,\omega_i)=\alpha(\omega_s+\omega_i)\phi(\omega_s,\omega_i)$, here: $$\alpha(\omega_s+\omega_i) = exp\left(-\frac{(\omega_s+\omega_i-\omega_{p0})^2}{\sigma_p^2}\right)$$ - pump envelope function, $$\phi(\omega_s,\omega_i)=\frac{1}{L}\int_0^Lg(z)exp(i\Delta k(\omega_s,\omega_i)z)dz$$ - phase matching function (PMF), $g(z)=\chi^{(2)}(z)/\chi_0^{(2)}$ is the normalised non-linear response of the material. For crystals with a uniform nonlinearity: $$\phi(\omega_s,\omega_i)=sinc\left(\frac{L\Delta k(\omega_s,\omega_i)}{2}\right)$$ (Note: if we consider integral in eq \eqref{eq:JSA} for $\xi<<1$ it will be proportional the same sinc function).

For achieving high purity it is necessary to remove spectral correlations of PMF with pump envelope function. The best way is to create a Gaussian-shaped PMF by adjusting the domain structure \cite{pickston2021optimised}. This shape allows us to achieve a single-mode PMF. Enter an objective function for nonlinearity modulation \cite{tambasco2016domain}:
$$g(z) = \exp{  \left(- \frac{\left( z - \frac{L}{2} \right)^2}{2 \sigma^2} \right)} \cos{\frac{2 \pi}{\Lambda} z}$$. 
Thus, after the Fourier transform g(z) - we get objective PMF:
\begin{eqnarray*}
    &&\phi \left( z, \Delta k = \frac{2\pi}{\Lambda} \right) = \\
    &&= \frac{1}{2L} \sqrt{\frac{\pi}{2}} \sigma \left( \text{erf} \left ( \frac{2z - L}{2 \sqrt{2} \sigma} \right) + \text{erf} \left ( \frac{L}{2 \sqrt{2} \sigma} \right) \right).
\end{eqnarray*}
Now, let's consider $g(z)$ which takes on discrete values of \textpm 1 for each domain and get PMF which 
Now, let's define the PMF with the condition that $g(z)$ takes on discrete values of \textpm 1 for each domain:
\begin{eqnarray*}
    &&\phi \left( z_m, \Delta k \right) = \frac{1}{L} \sum_{n=0}^{m} s_n \int_{n\omega}^{(n-1) \omega} \exp^{ -i k z} \,dz = \\
    &&=\frac{1}{L} \frac{\left( 1 - \exp^{ -i k \omega} \right)}{i k} \sum_{n=0}^{m} s_n \exp^{ -i k n \omega},
\end{eqnarray*}
where $\omega$ - is domain width, $s_n = $ \textpm 1 is the orientation of the n-th domain. Using the algorithm presented in \cite{graffitti2017pure}, it is possible to understand what the domain structure should look like in order for the PMF to have a Gaussian shape.

\begin{figure}[!h]
  \centering
  \includegraphics[width=0.99\linewidth]{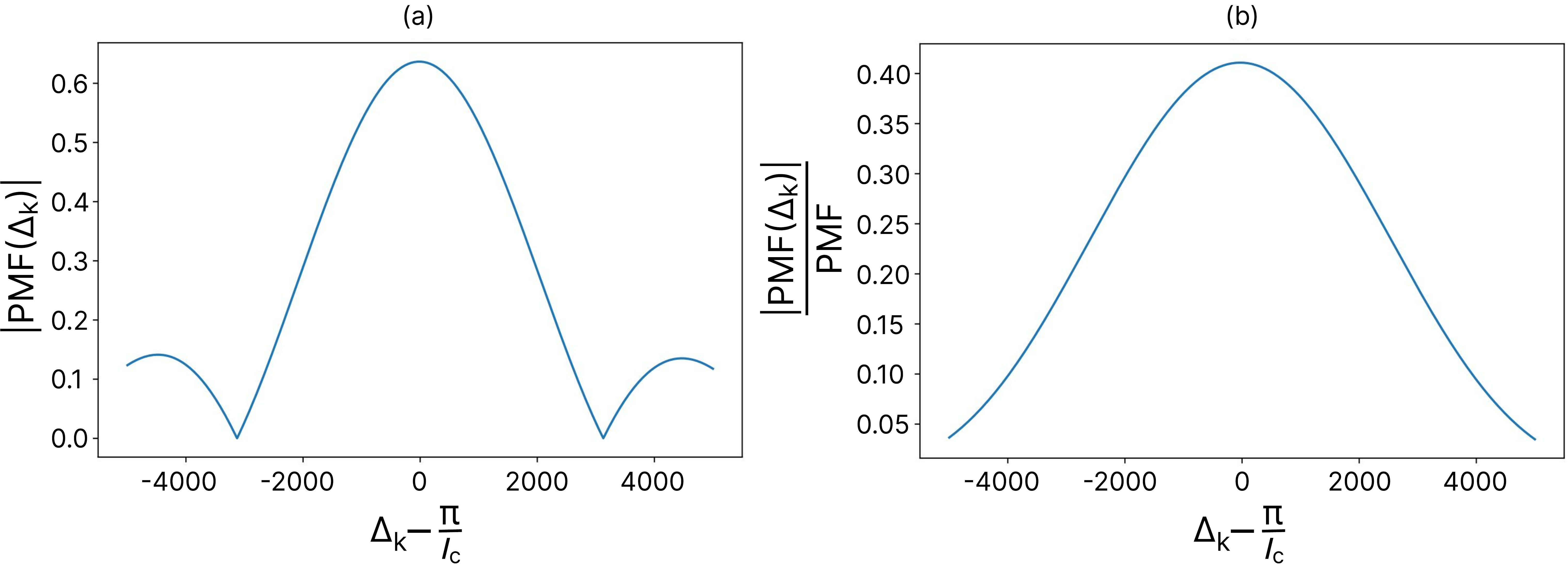},
  \caption{PMF for periodically poled and customized KTP crystal, PMF is normalized to the maximum value for the periodic structure. \label{fig:PMF for pp and ap}}
\end{figure}

On Fig. \ref{fig:PMF for pp and ap} compared PMF for periodically poled KTP crystal and result of numerical computations for periodic structure. In this way, the purity can be increased from 0.077 to 0.092.

Additionally, the purity of the emitted photons is influenced by the width of the pump beam spectrum and the length of the crystal. Thus, we are able to meticulously monitor how the purity varies in response to these variables in both our periodically polarized crystal and an aperiodic one, allowing us to elucidate the distinction between them with precision (see Fig. \ref{fig:Puriry from siglam length2}).

The purity of photon generation in a specific crystal is dependent on the wavelengths used, which is expected, given the phase mismatch condition. Consequently, for our chosen pump, signal and idler photon wavelengths, the generation efficiency cannot be significantly increased. For this reason, the sole method to enhance purity is to vary the spectral pump bandwidth and crystal length, as well as modify the domain configuration in order to mitigate undesired correlations.

In this instance, if we employ a crystal with a length of 2 millimeters, but with a specially designed domain structure, it is possible to enhance the purity by up to $10\%$. Based on the provided data, it can be concluded that in order to generate photons with reduced correlation in the spontaneous parametric downconversion (SPDC) process, lasers with a more narrow spectral width or pulse duration in the picosecond range should be utilized. And it is also worthwhile to increase the length of the crystal.

\end{document}